\documentclass[amsmath,amssymb,amsfonts,twocolumn,english,showpacs,superscriptaddress,prb]{revtex4-1}
\pdfoutput=1
\usepackage[T1]{fontenc}
\usepackage[latin9]{inputenc}
\usepackage{times}
\usepackage{graphicx}
\usepackage{color}
\usepackage{mathrsfs}
\usepackage{float}
\usepackage{indentfirst}
\usepackage{babel}

\usepackage[colorlinks,bookmarks=false,citecolor=blue,linkcolor=blue,urlcolor=blue]{hyperref}

\usepackage[sort&compress]{natbib}
\setcitestyle{numbers,square}

\usepackage{color}

\newcommand{\uone}{$U(1)$}

\newcommand{\eqnref}[1]{Eq.~(\ref{#1})}

\graphicspath{{../images/}{../diagrams/}{.}} 
\newcommand{\beq}{\begin{equation}}
\newcommand{\eeq}{\end{equation}}
\newcommand{\beqa}{\begin{eqnarray}}
\newcommand{\eeqa}{\end{eqnarray}}
\newcommand{\bi}{\begin{itemize}}
\newcommand{\ei}{\end{itemize}}
\newcommand{\ket} [1] {\vert #1 \rangle}

\newcommand{\ev}[1]{\langle #1 \rangle}

\newcommand{\vket}[1]{\left |#1 \right )}
 
\newcommand{\braopket}[3]{\left \langle #1 \middle |#2 \middle | #3 \right \rangle}

\newcommand{\I}{\mathcal{I}} 

\makeatother

\usepackage{wasysym}
\usepackage{booktabs}
\usepackage{subfigure}
\begin{document}

\title{Topological Crystalline Bose Insulator in Two Dimensions via  Entanglement Spectrum}
\author{Brayden Ware}
\affiliation{Department of Physics, University of California, Santa Barbara, CA 93106-6105, USA}

\author{Itamar Kimchi}
\affiliation{Department of Physics, University of California, Berkeley, CA 94720, USA}

\author{S. A. Parameswaran}
\affiliation{Department of Physics and Astronomy, University of California, Irvine, CA 92697, USA}
\affiliation{California Institute for Quantum Emulation, Elings Hall, University of California, Santa Barbara, CA 93106, USA}

\author{Bela Bauer}
\affiliation{Station Q, Microsoft Research, Santa Barbara, CA 93106-6105, USA}

\begin{abstract}
Strongly correlated analogues of topological insulators have been explored in systems with purely
on-site symmetries, such as time-reversal or charge conservation. Here, we use recently developed
tensor network tools to study a quantum state of interacting bosons which is featureless in the bulk,
but distinguished from an atomic insulator in that it exhibits entanglement which is protected by its
spatial symmetries.
These properties are encoded in a model many-body wavefunction that describes
a fully symmetric insulator of bosons on the honeycomb lattice at half filling per site.
While the resulting integer unit cell filling allows the state to bypass `no-go' theorems that trigger fractionalization at fractional filling, it nevertheless has nontrivial entanglement, protected by symmetry.
We demonstrate this by computing the boundary entanglement spectra, finding a gapless entanglement edge described by a conformal field theory as well as degeneracies protected by the non-trivial action of combined charge-conservation and spatial symmetries on the edge.
Here, the tight-binding representation of the space group symmetries plays a particular role in allowing
certain entanglement cuts that are not allowed on other lattices of the same symmetry,
suggesting that the lattice representation can serve as an additional symmetry ingredient in protecting
an interacting topological phase. Our results extend to a related insulating state of electrons, with short-ranged entanglement and no band insulator analogue.
\end{abstract}
\maketitle


\section{Introduction}

The experimental discovery~\cite{konig2007} of 
topological band insulators that cannot be adiabatically continued to the atomic limit as long as time-reversal invariance is preserved~\cite{Qi2011, Hasan2010} has spurred the exploration of a broad array of phases where symmetries protect subtle, non-local features that distinguish
them from trivial, unentangled insulators. These phases, collectively known as symmetry-protected topological (SPT) phases~\cite{chen2012}, have
by now been observed in several experimental realizations in one, two and three dimensions and an extensive
mathematical framework has been developed for their characterization and classification~\cite{Ryu2010, chen2011,schuch2011, Ryu2012, chen2013, wen2014, Wen2014-gv}.

Although a system that is in such a phase is often only subtly distinguished from a trivial insulator in its bulk, the topological properties can be observed by focusing on the boundary of a finite system. In many cases, the boundary exhibits gapless features such as the localized
Majorana zero modes at the ends of one-dimensional topological superconductors~\cite{kitaev2001},
the helical edge states of the quantum spin Hall effect~\cite{Haldane1988, Kane2005, Bernevig2006} and the protected Dirac cones on the surface of three-dimensional
$\mathbb{Z}_2$ topological insulators~\cite{FuKaneMele2007}.
It has been shown that these features are also manifest in the entanglement spectrum~\cite{li2008}, where they generally take the form
of either protected degeneracies or gapless spectra that mirror the gapless modes on a boundary.
These signify entanglement that cannot be removed while preserving the symmetry, and can thus be used to establish
that SPT phases are fundamentally distinct from trivial, non-entangled insulators.

Considering not only local symmetries but also symmetries that relate the physical locations of degrees of freedom,
such as the spatial symmetries of rotation or reflection, can lead to an even richer 
panoply  of phases~
\cite{Fu2011,Tanaka2012,Dziawa2012,berg2008,pollmannberg2012,Slager2013}
that require new probes as well as a more powerful classification.
The role of the boundary is also modified when spatial symmetries are considered: while in some cases, any possible physical boundary
of the system may break the relevant symmetry, the non-trivial features can still be extracted from the entanglement spectrum.
This allows the (numerical) identification of such SPT phases even when a necessary ingredient for the topological protection is a non-local symmetry operation.
However, while non-interacting topological crystalline insulators, i.e. band insulators whose non-trivial structure is protected
by crystal symmetries, have been well explored~\cite{Hughes2011, Turner2012}, far less is known about their interacting counterparts in two or three dimensions.

In this work, we compute the entanglement properties of an insulating state of interacting bosons on the honeycomb lattice,
and show that it constitutes a topological phase protected by lattice symmetries. In particular, we show that the non-trivial
entanglement is not related just to the group formed by the lattice symmetries, but becomes tied to the specific realization
as a honeycomb lattice. Combined with the symmetry, this enforces a non-trivial short-range entanglement structure.

The wavefunction we consider is an insulator of interacting bosons on the honeycomb lattice at a filling
of one boson per unit cell, or half filling per site. It forms one example of a class of insulators that
require a non-Bravais lattice, i.e. a lattice with multiple sites per unit cell.
The necessity for such a non-trivial unit cell arises due to higher-dimensional generalizations~\cite{Oshikawa:2000p1,PhysRevLett.90.236401,Hastings2004,Hastings:2005p1} of the Lieb-Schultz-Matthis (LSM) theorem~\cite{Lieb1961},
which forbids the existence of a featureless state
--- a state that neither spontaneously breaks a symmetry, nor displays intrinsic topological
order, nor has power-law correlations and is thus ``gapless'' --- in systems with a fractional filling per unit cell.
While such a featureless state at half-filling per site is allowed on the honeycomb lattice, the explicit
construction is challenging. Symmetry guarantees that a free-fermion spectrum is gapless at
certain high-symmetry points, and there is thus no basis of localized, symmetric and orthogonal Wannier states.
This implies that a featureless state on the honeycomb lattice cannot be constructed by filling a permanent of localized Wannier
orbitals~\cite{parameswaran2013}, and any construction of a quantum state thus must involve interactions.
Ref.~\onlinecite{kimchi2013} pursued an approach of constructing a permanent wavefunction by filling local and symmetric
orbitals that are not orthogonal and it was argued that that the resulting wavefunction is indeed featureless.
In particular, using numerical simulations it was found that the state exhibits isotropic and exponentially decaying correlations,
and arguments were presented that it is not topologically ordered.

Here while we confirm the featureless bulk of the state, including the absence of intrinsic topological order,
we show that nevertheless the entanglement of the state
cannot be removed while preserving
certain symmetries --- it is symmetry-protected. 
The relevant symmetry is a combination of charge
conservation and lattice symmetries, which together protect universal features in the entanglement spectrum. In
particular, we show that the low-lying entanglement spectrum is to great accuracy described by that of a $c=1$
conformal field theory, and that there is an exact double
degeneracy throughout the entanglement spectrum for certain geometries, which is protected
by the symmetries of the state and thus serves as a topological invariant identifying the SPT order.
Since lattice symmetry is involved crucially, this provides one of the first examples for an SPT of interacting bosons
protected by lattice rather than on-site symmetries.
To further substantiate the robustness of these features, we obtain parent Hamiltonians for the phase
in certain quasi-one-dimensional geometries and study the effect of weak symmetry-preserving and -breaking
perturbations on the ground states of these Hamiltonians.

All of these features 
become accessible through a description of the state
as a projected entangled-pair state (PEPS)~\cite{hieida1999numerical,nishino2001,gendiar2003,verstraete2004}. These states form a specific class of tensor
network states that corresponds to a generalization of the well-known matrix-product state
(MPS)~\cite{white1992,fannes1992,ostlund1995, schollwock2011}
framework to higher dimensions. PEPS have been shown to be a powerful description of many
classes of gapped systems, including topologically ordered and SPT phases. Here, we have an
exact description of the state as a PEPS, allowing us to extract properties such as the entanglement
spectrum and the topological invariants exactly on certain geometries; we emphasize that these
properties of the state are not accessible to other numerical methods.

The topological invariants extracted here form examples of a broad class
of invariants that provably must be constant throughout the phase. These differ
from the order parameters that measure local symmetry breaking in that they
are not related to the expectation values of local operators. Early examples
of topological invariants for SPT phases are the string order parameter for the one-dimensional
AKLT phase~\cite{affleck1987, affleck1988, denNijs1989, pollmannberg2012},
and the spin Chern number for the quantum spin Hall effect~\cite{sheng2006}.
The invariants we consider here measure how the action of the symmetry is implemented on the
physical edge states of open systems or on the Schmidt states of an
entanglement decomposition~\cite{pollmann2010}. These invariants feature heavily in the
classification of SPT phases with on-site symmetry, and similar invariants
that apply to free-fermion states have been used for topological crystalline
insulators~\cite{Fu2007, Hughes2011}. In contrast, topological invariants for interacting states
protected by lattice symmetries in more than one dimension are poorly 
understood. We will discuss the action of the symmetry on the edge of the
state and progress towards the goal of finding a topological invariant to
identify the corresponding phase.

The rest of this paper is structured as follows: in Section~\ref{sec:fbi}, we review the
honeycomb featureless boson insulator (HFBI) state, and introduce its PEPS representation.
In Section~\ref{sec:Correlations}, we discuss results for the correlation functions
of this state. In Section~\ref{sec:ES}, we discuss the entanglement spectra that
we obtain numerically from the PEPS representation and discuss in detail
their connection to the spectrum of a free boson conformal field theory.
In Section~\ref{sec:symmetry}, we describe the symmetry-protected topological
invariants that allow us to identify the symmetries that protect certain entanglement
properties of the state in quasi-one-dimensional geometries. Section~\ref{sec:perturbations}
discusses the effect of weak perturbations to a parent Hamiltonian in the same
quasi-one-dimensional setup, and Section~\ref{sec:duality} introduces a different
perspective on the phase from the point of view of a boson-vortex duality.


\section{Construction of the featureless boson insulator}
\label{sec:fbi}

In the honeycomb lattice, each unit cell is associated with exactly one hexagon plaquette, which respects the lattice point group symmetries. As shown in Ref.~\onlinecite{kimchi2013}, this provides an explicit
construction of a bosonic insulator
on the honeycomb lattice that is completely featureless in the bulk,
henceforth referred to as \emph{honeycomb featureless boson insulator} (HFBI).
The state is succinctly described by the following  expression:
\begin{equation} \label{eq:def}
\ket{\psi} = \prod\limits_{\hexagon} \left( \sum\limits_{i \in \hexagon} b^{\dagger}_{i} \right) \ket{0}.
\end{equation}
Here, $\hexagon$ denotes the elementary hexagons of the honeycomb
lattice. Despite the deceivingly compact expression,
this many-body bosonic state is strongly correlated
and requires concrete computation for its properties to be unveiled.

We focus on two closely related variants of this state: a
version of soft-core bosons where $b_i^\dagger$ creates a boson on
site $i$ and obeys the usual bosonic commutation relations, and a
hard-core version of the same state where $b_i^\dagger$ also creates a
boson but $(b_i^\dagger)^2=0$. In either case, the operator $\sum_{i
\in \hexagon} b^{\dagger}_{i}$ creates exactly one boson per
hexagon; as there is one hexagon per unit cell of two sites of the
lattice, the state has one boson per unit cell, or half a boson per
site, thus allowing the existence of a featureless state.
In the case of soft-core bosons, the maximum number of bosons
per site is three.

Ref.~\onlinecite{kimchi2013} examined properties of both the soft-core and hard-core variants of this state. In the soft-core case, ground-state correlations were mapped to those of a classical loop model on the triangular lattice. A Monte Carlo analysis thereof revealed that the boson Green's function $\langle b^\dagger_i b^{\phantom{\dagger}}_j\rangle$ decays exponentially -- thereby ruling out the possibility that the many-body wavefunction describes a superfluid -- and further, that the correlation functions of a variety of neutral operators (e.g., those describing charge- or bond-density order) remained short-ranged. This loop model mapping also included a variational parameter, $m$, that tunes the the soft-core boson wavefunction on the honeycomb lattice into that of a trivial Mott insulator on a triangular lattice of fictitious sites placed at the center of each hexagon. The absence of a transition under this perturbation was taken as evidence that the ground state would remain unique on manifolds of nontrivial topology, thereby ruling out the possibility that the wavefunction describes a topologically ordered phase. For the case of hard-core bosons, a different quantum-classical mapping enabled the efficient calculation of boson number correlations. This directly revealed that the hard-core projection did not induce any long-range correlations in the neutral sector. Although working in the number basis precludes direct access to `charged' correlators such as the boson Green's function, on general grounds, the algebraic decay characteristic of classical 2D superfluids~\footnote{Note that the quantum-classical mappings described map quantum correlations to classical ones {\it in the same dimension}.} is expected to also infect density-density correlations, and thus their exponential decay provides indirect evidence that the hard-core boson wavefunction also lacks superfluid order.
We note that none of these quantum-classical mappings can readily provide insights into the entanglement properties of the wavefunction.

\subsection{PEPS representation}

In order to go beyond the properties accessible via these quantum-classical
mappings of the HFBI, and in particular in order to be able to study its edge
properties, we now derive a representation as a projected entangled
pair state (PEPS). Importantly, this PEPS description will respect
all of the relevant symmetries of \eqnref{eq:def}.

\begin{figure}
	\centering
	\includegraphics[width=0.8\columnwidth]{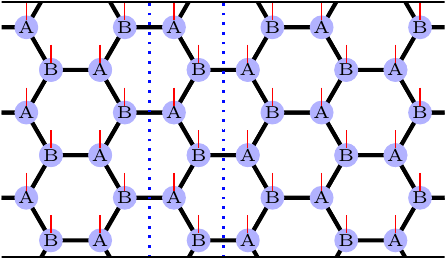}
	\caption{Honeycomb lattice PEPS and zig-zag entanglement cut.
	In this PEPS of rank-4 tensors, the top and bottom edges are identified, forming a cylinder with circumference $W=3$ unit cells. 	
	A one-dimensional MPS representation is constructed by contracting the tensors in each cylinder slice (region marked by dotted lines).
	The entanglement cut used (either one of the dotted lines) passes through the hexagon mid-points, preventing the tight-binding lattice from gaining additional sites as long as crystalline symmetries are preserved.
	}
	\label{fig:PEPS}
\end{figure}

\begin{figure}
	\centering
	\includegraphics[width=0.8\columnwidth]{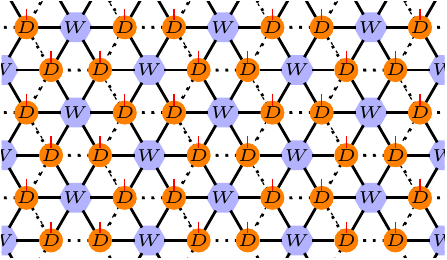}
	\caption{
	Intermediate tensor network for HFBI state. Here, the tensors labeled
	$D$ are located on the sites of the honeycomb lattice, while the
	tensors labeled $W$ are located on the centers of each hexagon.
	Dotted lines thus represent the physical lattice, while the solid
	lines indicate auxiliary bonds over which the tensor network is
	contracted.}
	\label{fig:FBI_PEPS}
\end{figure}

To obtain a PEPS construction, we first choose a local basis $\ket{n}$
of boson occupation numbers, i.e. $b^\dagger b \ket{n} = n \ket{n}$.
The PEPS will thus describe the coefficients of $\ket{\psi}$ in this
basis, $\langle n_1 \ldots n_L | \psi \rangle$. The PEPS
representation is most easily obtained in a two-step construction,
where we first construct the state shown in Fig.~\ref{fig:FBI_PEPS}.
Here, the tensor labeled $W=W^{n_1 \ldots n_6}$, which is placed in
the center of each hexagon, is a rank-6 tensor given by
\begin{equation} \label{eq:W}
W^{n_1 \ldots n_6} = \left\{ \begin{array}{ll}
													1 : & \sum\limits_i n_i = 1 \\
													0 : & \text{else}
													\end{array} \right.,
\end{equation}
where each $n_i \in \{0, 1\}$.

This tensor describes the coefficients of a so-called $W$-state in the
occupation number basis, i.e. $W^{n_1 \ldots n_6}= \langle n_1
\ldots n_6 | \sum_{i=1}^6 b_i^\dagger |0\rangle$. We note that this
tensor is symmetric under permutations of its indices.

\begin{figure}
	\centering
	\subfigure[D tensor]{%
		\includegraphics[width=0.18\columnwidth]{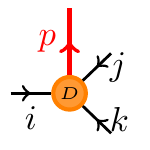}
		\label{fig:D}
	}
	\quad
	\subfigure[W-tensor and factored form]{%
		\includegraphics[width=0.5\columnwidth]{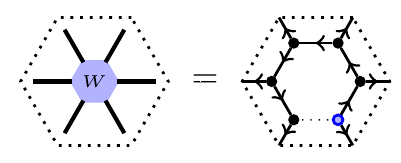}
		\label{fig:W}
	}
	\qquad
	\subfigure[A tensor]{%
		\includegraphics[width=0.2\columnwidth]{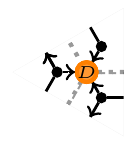}
		\label{fig:A}
		}
	\quad
  \subfigure[B tensor]{%
		\includegraphics[width=0.2\columnwidth]{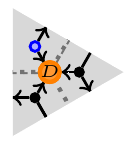}
		\label{fig:B}
		}
	\subfigure[PEPS tensor network for F.B.I. state]{%
		\includegraphics[width=0.8\columnwidth]{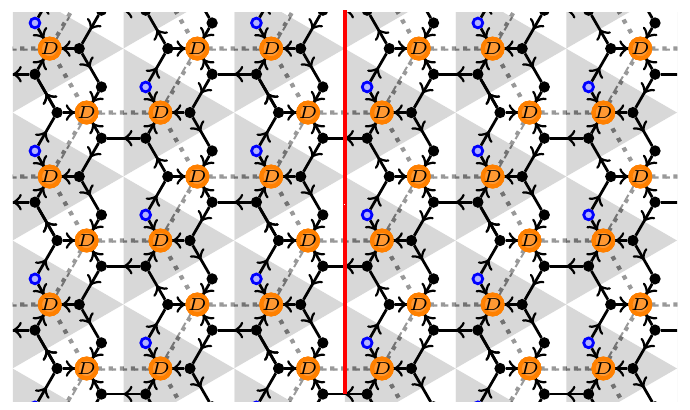}
		\label{fig:FBI_PEPS_2}
	}
\caption{Construction of PEPS for HFBI state.
The site tensors (shown in panels (c), (d)) are constructed
using the factors of the plaquette tensor $W$ (panel (b))
combined with the original vertex tensor $D$ (panel (a)).
The red line in panel (e) shows where the entanglement cut considered in this paper
crosses the network.
}
\label{fig:PEPSforFBI}
\end{figure}

On the sites of the physical lattice, we have placed a rank-4 tensor
denoted as $D$, shown in panel (a) of Fig.~\ref{fig:D}, which
connects the $W$ tensors from three adjacent hexagons, and as fourth
index has a physical index $p$. For a state of soft-core bosons, where
$p=0,1,2,3$, this tensor is given by
\begin{equation} \label{eqn:Dsc}
D^\mathrm{sc}_{p, i_0 i_1 i_2} = \left\{ \begin{array}{ll}
													\sqrt{p!}\hphantom{0} : &  p=i_0+i_1+i_2 \\
													0\hphantom{\sqrt{p!}} : &  \text{else}
											\end{array} \right..
\end{equation}

We can also encode a state of hard-core bosons by replacing $D$ by
\begin{equation} \label{eqn:Dhc}
D^\mathrm{hc}_{p, i_0 i_1 i_2} = \left\{ \begin{array}{ll}
													1 : & p = i_0+i_1+i_2 \le 1 \\
													0 : & \text{else}
													\end{array}
											\right..
\end{equation}
Other values for the $D$ and $W$ tensors that respect the charge and lattice symmetries
can also give rise to featureless insulators.
Some of these variants are described in Appendix~\ref{Appendix:Variants}.

This tensor network wavefunction manifestly respects all the
translational and point group symmetries of the honeycomb lattice,
since the tensors $W$ and $D$ are invariant under rotations of their
virtual indices in the plane. One can also check that the
wavefunction is \uone{} invariant with charge $1$ per plaquette.

In order to convert the tensor network of Fig.~\ref{fig:FBI_PEPS} into a PEPS representation,
we first factor the $W$-tensor into a matrix-product state
of six tensors as shown in Fig.~\ref{fig:W}. We
choose a form of the MPS that breaks the rotational symmetry
of the W-state (which appears as translational symmetry of the MPS).
This allows us to obtain an MPS description with a small bond dimension
of $M=2$; a fully symmetric choice would require bond dimension 6.
Since these states are physically equivalent, all
physical quantities are unaffected by this choice.
One possible decomposition is given by
\begin{equation}
W^{i_1 i_2 i_3 i_4 i_5 i_6} = \sum\limits_{\alpha_1 \ldots \alpha_5} V^{i_1}_{\alpha_1} W^{i_2}_{\alpha_1 \alpha_2}
\ldots
W^{i_5}_{\alpha_4 \alpha_5} X^{i_6}_{\alpha_5}
\end{equation}
where $V^{i_1}_{\alpha_1} = \delta_{i_1, \alpha_1}$, $X^{i_6}_{\alpha_5} = \delta_{i_6,\alpha_5+1}+\delta_{i_6,\alpha_5-1}$, and
\begin{equation*}
W_{i_0 i_1}^{j} = \begin{cases}
													1 &: i_0+j=i_1 \\
													0 &: \text{else}
											\end{cases},
\end{equation*}
where each index takes values in $\{0, 1\}$. Applying this to each
$W$-tensor yields the state as shown in Fig.~\ref{fig:FBI_PEPS_2}.
By contracting the four tensors in each
shaded region together, we obtain a PEPS in the regular form as shown
in Fig.~\ref{fig:PEPS}. The resulting PEPS has a bond dimension of
$M=2$ on the horizontal bonds, and a bond dimension of $M=4$ on all
other bonds. While it superficially breaks
the rotational symmetry of the lattice, it is an exact representation
of the FBI state and does not break any symmetries after contracting
the indices.

This decomposition respects the physical \uone{} charge conservation
symmetry in that all tensors are separately
\uone{}-invariant~\cite{bauer2011}. To make this manifest, we have
indicated in Fig.~\ref{fig:PEPSforFBI} arrows on each bond that show
the flow of charge.

\subsection{Representation on infinite cylinders}
For the calculations presented in this manuscript, we consider the
state $\ket{\psi}$ on a cylinder of infinite length, but finite
circumference of $W$ unit cells. In Fig.~\ref{fig:PEPS}, we have indicated the
choice of boundary conditions for the cylinder used in this paper.
For many practical purposes, the PEPS on an infinite cylinder can be represented as an infinite, \
translationally invariant matrix-product state of bond dimension $\chi=2^W$
and physical dimension $p=4^{2W}$ ($p=2^{2W}$) for the soft-core (hard-core) case.
The MPS is created by blocking all tensors in each slice of the cylinder, as shown in
Figure~\ref{fig:PEPS}.

With each cylinder slice blocked together and considered as an MPS,
the procedures we use for computing both correlation functions and entanglement properties
are in principle identical to those used previously in MPS \cite{cirac2011,pollmann2010}.
Due to the exponential increase in the MPS bond dimension, this numerically exact approach
scales exponentially in the circumference of the cylinder.
It is however computationally advantageous to exploit the additional
structure present in the PEPS transfer operator;
by doing so, we can compute correlation functions and
the entanglement spectrum for the cut shown in Figure \ref{fig:FBI_PEPS_2}
for the HFBI state on cylinders of circumference up to $W=10$.
These computations are presented in the following
Sections~\ref{sec:Correlations} and \ref{sec:ES}.


\section{Correlation functions}
\label{sec:Correlations}

In Ref.~\onlinecite{kimchi2013}, certain real-space correlation functions
of the featureless boson insulator state were studied using a mapping to
a particular classical statistical mechanical system which was sampled using
Monte Carlo techniques.
Here, we go beyond this by employing PEPS calculations on infinite cylinders
that allow us to measure a broader class of correlation functions and, in
particular, allow us establish a strict upper bound on the exponential decay
of \emph{all} two-point correlation functions for an infinite cylinder of
given width.

\begin{figure}
	\centering
	\includegraphics[width=\columnwidth]{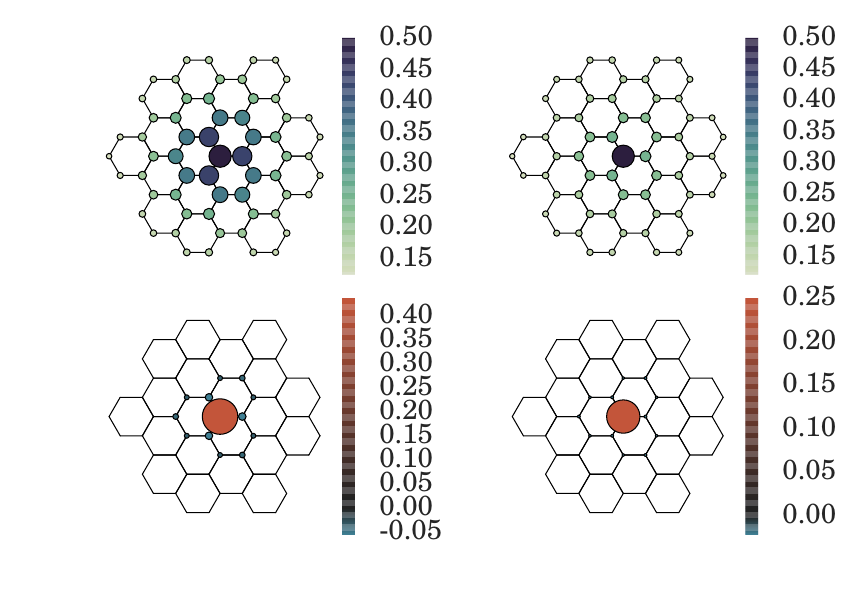}
	\vskip-5ex
	\caption{
	 Short distance correlation functions $\ev{b_0^{\dagger} b_x}$ (top panels)
	 and $\ev{(n_0-\frac12)(n_x-\frac12)}$ (bottom panels)
	 for the soft-core HFBI (left panels) and the hard-core HFBI (right panels)
	 on the $W=8$ cylinder.
	 The magnitude of the correlation function at site $x$ is proportional to
	 the radius of the corresponding circle.
	 }
	\label{fig:ShortCorr}
\end{figure}

In Fig.~\ref{fig:ShortCorr}, we show both density-density and off-diagonal
short-range correlation functions for a cylinder of circumference $W=8$.
Comparing these to the Monte Carlo results of Ref.~\onlinecite{kimchi2013},
which have been computed for a different geometry, we find good qualitative
agreement. Crucially, while the boundary conditions we choose break the rotational
symmetry by making the system periodic in one direction and infinite in the other,
the short-range correlations for distances up to half of the cylinder circumference
appear unaffected by this.

It is a well-known result that PEPS can, in the thermodynamic limit,
exhibit power-law correlation functions~\cite{verstraete2006}, while the correlation
functions in an MPS of finite bond dimension decay exponentially. The
long-range correlations of an MPS are encoded in its transfer operator $T$,
which for an MPS of bond dimension $M$ is a matrix of size $M^2 \times M^2$.
Denoting the spectrum of $T$ as $\lambda_i$ with $|\lambda_0| \geq |\lambda_1| \geq \ldots$,
we can normalize the state such that $\lambda_0 = 1$. If the largest eigenvalue
is found to be non-degenerate, $\lambda_1 < \lambda_0$, we have that all correlation
functions of operators $\mathcal{O}_i$ that are supported on a finite number of sites centered
around a site $i$ decay as
$\langle \mathcal{O}_i \mathcal{O}_j \rangle - \langle \mathcal{O}_i \rangle \langle \mathcal{O}_j \rangle \sim e^{-|i-j|/\xi_\mathcal{O}}$.
Crucially, the correlation length $\xi_\mathcal{O}$ for any operator $\mathcal{O}$
is bounded from above by $-1/\log |\lambda_1|$~\cite{schollwock2005}.
In the following, we thus
evaluate the spectrum of the transfer operator of our PEPS along cylinders of varying
circumference $W$ to establish an upper bound on the correlation length for each
circumference $\xi(W)$
Note that the possibility of having power-law correlations in a PEPS can
be reconciled with the above consideration if the correlation length $\xi(W)$
diverges as $W \rightarrow \infty$; we will thus need to carefully consider the scaling
of $\xi(W)$.

\begin{figure}
	\centering
	\includegraphics[width=\columnwidth]{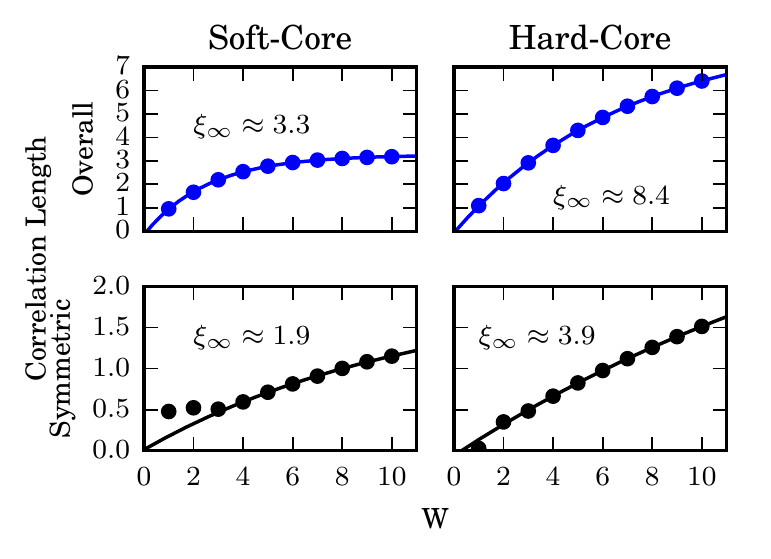}
	\vskip-2em
	\caption{Bound on the correlation length of all operators and \uone{} symmetric operators
	 respectively for the soft-core and hard-core states vs. cylinder circumference $W$.
	 Fits of the form $\xi = \xi_{\infty} - Ae^{-W/B}$ were
	 used to extract the correlation lengths.
	 These bounds can be confirmed to match the correlation lengths of
	 $\ev{b^{\dagger}_x b_y}$ and $\ev{n_x n_y}$ in each case.
	}
	\label{fig:TMS}
\end{figure}

Our results for the correlation bounds $\xi(W)$ are shown in Fig.~\ref{fig:TMS}.
Here, we show the upper bound for the case of soft-core and hard-core bosons,
and in each case consider the spectrum of the full transfer operator as well
as $S^z=0$ sector, which encodes correlations of operators $\mathcal{O}_i$ that do not
change the boson number (such as density-density correlations).
We find that in each case, the largest eigenvalue of the transfer operator is
non-degenerate. Furthermore, we find
that the correlation length approaches a finite constant as we
increase $W$, as shown in Figure \ref{fig:TMS}.


\newcommand{\uL}{\mathbf{L_0}}
\newcommand{\bL}{\mathbf{\bar{L}_0}}

\section{Entanglement spectrum}
\label{sec:ES}

The quasi-1D cylinder geometry is convenient for calculating the
entanglement spectrum for entanglement cuts transverse to the long direction of the
cylinder. Here, the entanglement spectrum $\varepsilon_i$ is defined through the spectrum
$\rho_i$ of the reduced density matrices $\rho_{L/R}$ obtained for a bipartition of the state,
where we have $\varepsilon_i = -\log \rho_i$.
The corresponding eigenvectors of the reduced density matrices
are referred to as Schmidt states $\ket{\psi^{(i)}_{L/R}}$.
The Schmidt decomposition
\beq
\ket{\psi} = \sum\limits_i \sqrt{\rho_i}
\ket{\psi^{(i)}_{L}}\ket{\psi^{(i)}_{R}}
\label{eq:schmidt}
\eeq
relates the Schmidt states to the original wavefunction and is useful to keep in mind
when interpreting the entanglement spectrum.

To extract the entanglement spectrum exactly,
we use a method proposed in Ref.~\onlinecite{cirac2011}.
In the setup given here, the exact representation of the HFBI state as a PEPS of
fixed bond dimension implies an upper bound on the number of non-zero $\rho_i$;
for the cut shown in Fig.~\ref{fig:FBI_PEPS_2}, this upper bound is $\chi=2^W$.

Upon computation of the entanglement spectrum, we find that this bound is saturated,
so that there are precisely $2^W$ contributing terms in Eq.~\ref{eq:schmidt}.
This fact can be simply understood without reference to the PEPS representation:
Each of the $w$ plaquettes on the cut can contribute its one boson either to the left
or the right of the cut, and this is the complete source of the uncertainty
of the state on one half of the cut when ignoring the state on the other half.

We can form a set of $2^W$ vectors $|\sigma_1,\ldots,\sigma_W)$ that correspond to the
choices for the auxiliary degrees of freedom of the PEPS across the cut, where
$\sigma_i \in \{0, 1\}$ is the number of bosons contributed by the $i$'th hexagon
the left of the cut.
The PEPS defines
a map from these boundary vectors to physical states in the bulk of the
semi-infinite cylinder, which we denote as $\ket{\psi_L^{(\sigma_1, \ldots, \sigma_W)}}$;
on the subspace of physical states spanned by Schmidt states with non-vanishing contribution
to the reduced density matrix, this map is invertible and can be computed explicitly.

Translation around the cylinder acts in the natural way on the states
$\ket{\psi_L^{(\sigma_1, \ldots, \sigma_W)}}$ by permuting the values of
$\sigma_i$, $\sigma_i \to \sigma_{i+1}$.
Although the boson number on the half-infinite cylinder is infinite, we can define for
each basis element a \uone{} charge corresponding to the number of bosons to the
left of the cut relative to a uniform background charge,
\begin{equation}
Q_L = \sum_i (\sigma_i - \frac12).
\end{equation}
Only relative charges between states will be important for our conclusions, and the precise
way in which background charge is accounted for does not matter.
Each state is paired with a corresponding state on the right half of the cylinder with opposite charge.

We can block-diagonalize the reduced density matrix
for a half-infinite cylinder in both the \uone{} charge and transverse momentum
quantum numbers, allowing us to perform more efficient calculations.
In addition, we can assign quantum numbers to both the Schmidt states and the entanglement
spectrum. This property is generically true for \uone{}-symmetric and translationally invariant PEPS
on a cylinder, although not in general true for arbitrary symmetry groups.
This point is elaborated on in Appendix~\ref{Appendix:MPS}.

\begin{figure}
	\centering
	\includegraphics[width=\columnwidth]{{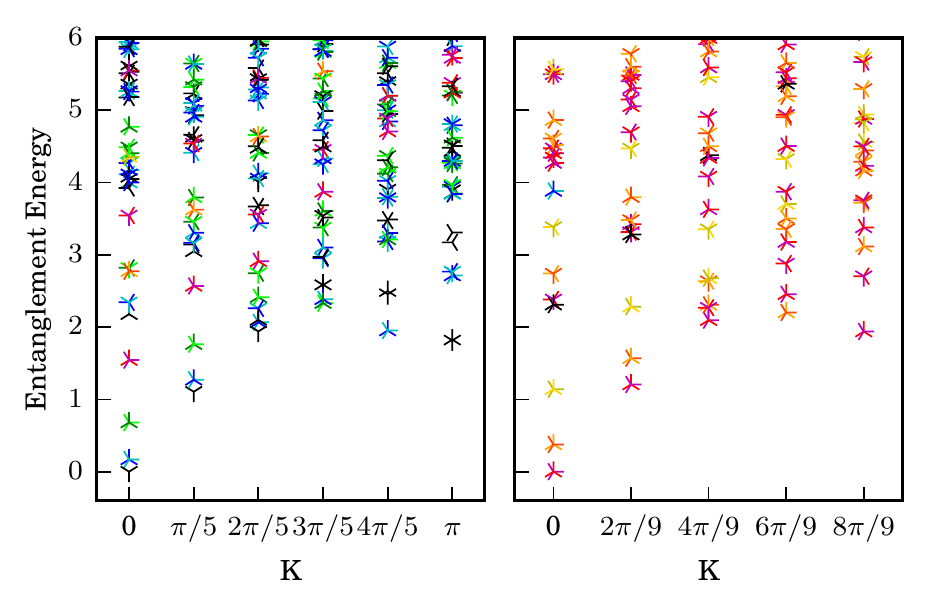}}
	\vskip-3ex
	\caption{
    Entanglement spectrum on the zig-zag edge of a cylinder of
    circumference $W=10$ (left panel) and $W=9$ (right panel),
    as function of transverse momentum $K_y$.
    Different colors indicate different \uone{} charge sectors. }
	\label{fig:ESL910}
\end{figure}

\begin{figure}[htbc]
	\centering
	\includegraphics[width=\columnwidth]{{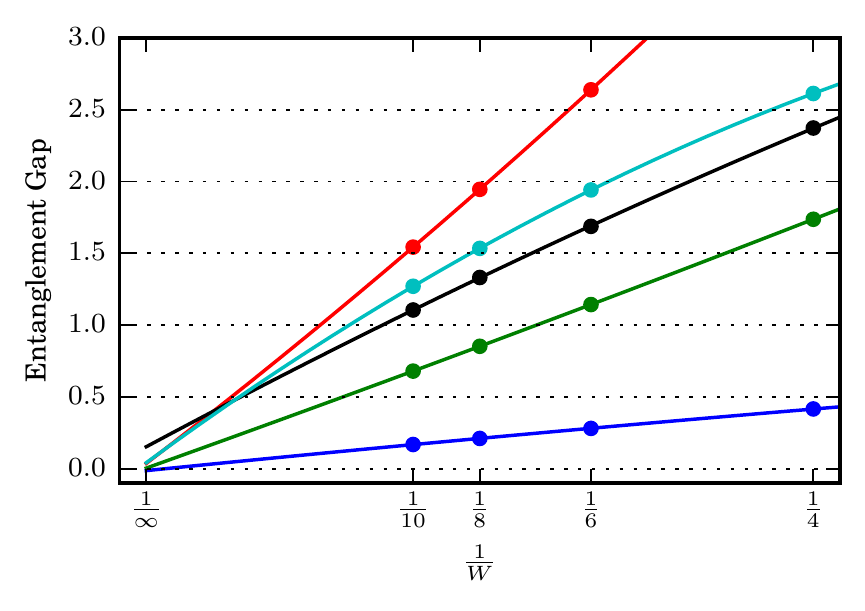}}
  \vskip-3ex
  \caption{
   The lowest five states above the ground state in Figure~\ref{fig:ESL910} show gapless $1/W$ scaling behavior. In this plot, fits for the entanglement energy versus $1/W$ (of the form $a +  \frac{b}{W} + \frac{c}{W^2}$) to extract the gap are consistent with a gap value of 0.}
  \label{fig:EEScaling}
\end{figure}

The entanglement spectra for the HFBI on cylinders with even and odd width
circumferences are shown in Fig.~\ref{fig:ESL910}, plotted
against the transverse momentum eigenvalue and colored by the \uone{} charge
eigenvalue of the corresponding Schmidt states. All the numerical results in this section
are obtained for the soft-core boson variant of the state.
We find that the entanglement spectrum looks like it has a gapless
edge mode with linear dispersion near momentum zero.
To further substantiate this, we compare the lowest entanglement energies
for several cylinder widths and quantum number sectors in Fig.~\ref{fig:EEScaling}.
The finite-size scaling confirms in all cases that
the entanglement gap closes as $1/W$, as you would expect for a gapless mode
with linear dispersion.

\begin{figure}[htbc]
	\centering
  \includegraphics[width=\columnwidth]{{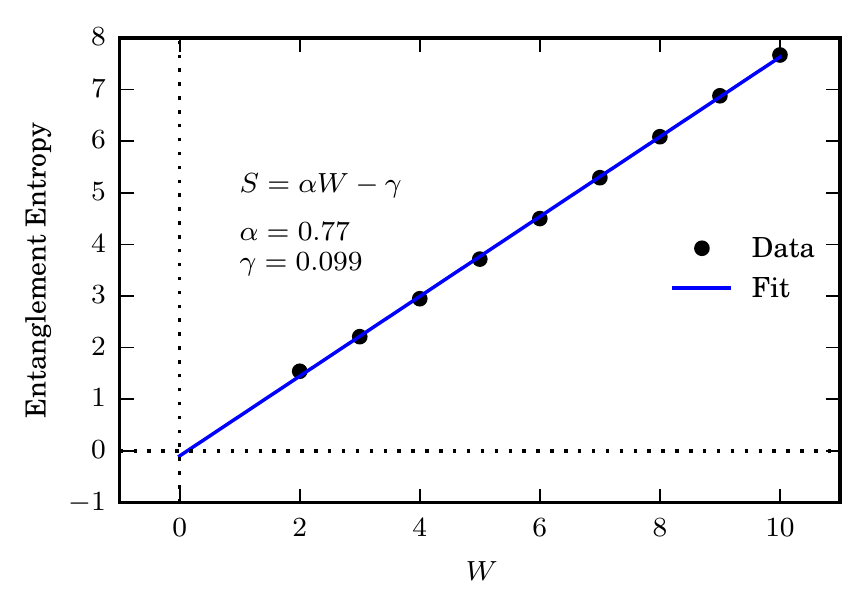}}
	\vskip-3ex
	\caption{The constant correction to the entanglement entropy, which measures the topological entanglement entropy $\gamma$ when the state is minimally entangled, is consistent
	with 0.}

	\label{fig:TopologicalEE}
\end{figure}

The gapless edge is suggestive of the state having either topological or SPT order.
While topological order was already ruled out in Ref.~\cite{kimchi2013}, our PEPS
representation gives us additional tools to substantiate this assertion. In particular
we can calculate the entanglement between the two parts of the cylinder as a function
of circumference $S(W)$ and check for a subleading term to the area law by performing
a fit to $S(W) = \alpha W + S_0$. In a topological phase and in one of the minimally entangled
states (MES)~\cite{zhang2012}, one would expect the subleading term to correspond to the topological entanglement
entropy, $S_0 = -\gamma$~\cite{kitaev2006, levin2006, jiang2012}.
In a non-minimally entangled state, one would instead measure other values of $S_0 > -\gamma$.
However, since each MES exhibits long-range order of a specific Wilson loop operator, such a
superposition of MES represents a superposition of different ordering patterns and can thus
be detected via a degeneracy of the largest eigenvalue of the transfer matrix.
Our results for the entropy are shown in Fig.~\ref{fig:TopologicalEE}. We find results that are consistent
with $S_0 = 0$, which together with the fact that we also find that the transfer matrix to be nondegenerate
rules out topological order.

\subsection{Conformal field theory description of the edge}
\label{sec:CFT}

In addition to the gapless behavior, we find that the low energy entanglement spectrum
can be completely described by the finite-size spectrum of a conformal field theory (CFT).
Given the \uone{} symmetry of the state, the simplest possible
conformal field theory we might expect to appear at the edge is that
of a single free bosonic field - and indeed, this is the CFT that matches the
entanglement spectrum.
We briefly review the relevant properties of this CFT~\cite{difrancesco}.

The free boson CFT is created from the Lagrangian
\begin{equation}
\mathfrak{L} = \frac{g}{2}\int dt \int\limits_0^W dx \left[ (\partial_t \phi)^2 - (\partial_x \phi)^2 \right]
\end{equation}
with the compactified field identification
\begin{equation*}
\phi \equiv \phi + 2\pi R
\end{equation*}
and placed on a circle of circumference $W$ with periodic boundary conditions
\begin{equation*}
\phi(x) \equiv \phi(x+W).
\end{equation*}
The family of free-boson CFTs is parametrized by a single parameter
$\kappa = \pi g R^2$, also known as the Luttinger liquid parameter~\cite{difrancesco, giamarchi}.

Upon canonical quantization, we find that the set of energy
eigenstates consists of \uone{} Kac-Moody primaries $\ket{e, m}$, with
integers $e, m$ labeling the \uone{} charge and the winding number of
the bosonic field respectively, and level $n, \bar{n}$ descendants of each primary for
non-negative integers $n,\bar{n}$,
which we will collectively label $\ket{e, m; n, \bar{n}}$.
The number of level $(n,\bar{n})$ descendants of a given
primary, all of which are degenerate in the thermodynamic limit, is $Z(n) Z(\bar{n})$, where
$Z(n)$ is the number of partitions of the integer $n$.

The energies and momenta for the states $\ket{e, m; n, \bar{n}}$ are given below
on a finite size system of circumference $W$:
\beq
\label{eq:finitesizespec}
\begin{split}
	\mathbf{P} &= \frac{2\pi}{W}(em + n - \bar{n}) \\
	\mathbf{H} &= \frac{2\pi}{W}(\frac{e^2}{4\kappa} + \kappa m^2 + n + \bar{n}) + \ldots 
\end{split}
\eeq
Here, the ellipsis ($\ldots$) denotes further subleading contributions due to coupling to irrelevant operators.

By rescaling the energy and momentum, we find a system-size
independent pattern that can be matched to the low-energy, linearly
dispersing part of the entanglement spectrum from Figure~\ref{fig:ESL910}:
\beq
\label{eq:finitesizespecscaled}
\begin{split}
\mathbf{P} &\propto (em + n - \bar{n})  \\
\mathbf{H} &\propto e^2 + 4\kappa^2 m^2 + 4\kappa(n + \bar{n}) + \ldots
\end{split}
\eeq

\begin{figure}[htbc]
	\centering
	\includegraphics[width=\columnwidth]{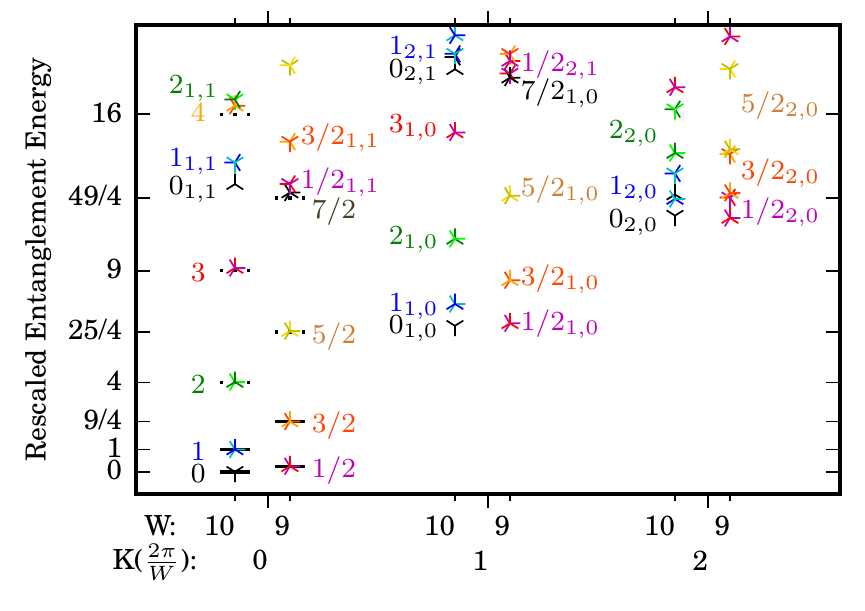}
		\vskip-3ex
	\caption{The identification of the primary states $\ket{\pm e, m=0}$ and the level $n, \bar{n}$
	descendants in the spectrum of the soft-core boson entanglement Hamiltonian. The states are labeled
	$e_{n, \bar{n}}$. The zero and scale of the numerical spectrum are set by matching the lowest two
	states. The energies and charges of the primaries with charges $2, 5/2, ... 4$ appear at the predicted
	spots.  The best estimate for the Luttinger parameter from this spectrum is $\kappa \approx 1.6$,
	taken from the energy of the $0_{1, 0}$ state. }
	\label{fig:EEIdentify}
\end{figure}

The results of this match are shown in Figure~\ref{fig:EEIdentify}.
An estimate for $\kappa$ can be obtained from the energy of the first descendent
$\ket{0, 0; 1, 0}$, which gives $\kappa \approx 1.6$. The label $e$, which measures the \uone{}
charge, is integer for even $W$ and half-integer for odd $W$. The degeneracy pattern $1, 1, 2, ...$ for the level-$(n,0)$ descendents along the edge of the cone matches the prediction.

\begin{figure}[htbc]
	\centering
	\includegraphics[width=\columnwidth]{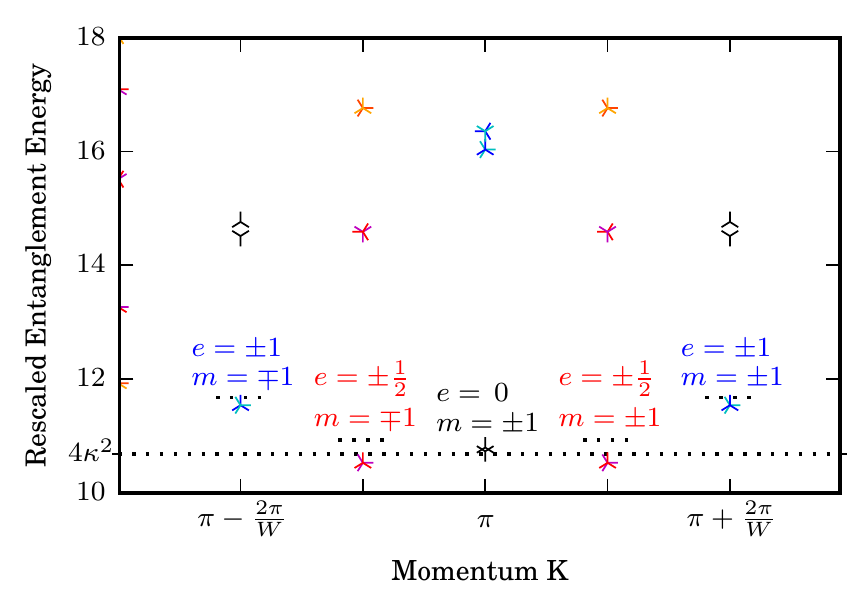}
		\vskip-3ex
	\caption{The identification of primary states $\ket{e, m=\pm1}$ and first descendants in the low energy part of the spectrum near momentum $\pi$. Unlike the $m=0$ states shown in Figure~\ref{fig:EEIdentify}, these primary states have shifted momentum $K = \pi + em (2\pi/W)$ and an extra double degeneracy due to the two values of $m=\pm1$. Using the estimate $\kappa \approx 1.6$ from Figure~\ref{fig:EEIdentify}, the predicted value of the entanglement energy for the $\ket{e=0, m=\pm1}$ states is $4\kappa^2$, which has been marked in the plot. The agreement is very good.}
	\label{fig:EEIdentifyPi}
\end{figure}

The states with
odd  winding number $m$, such as $\ket{0, \pm 1}$, do not appear at
the energy and momentum predicted by the above formula. Instead, the primary states $\ket{e, m=\pm
1}$ can be found centered around momentum $K=\pi$.
The identification of these states in the spectrum
is shown in Figure~\ref{fig:EEIdentifyPi}. Although the larger-$m$ states are too high in energy
to be reliably distinguished at this system size, a natural conjecture is that all primaries with
odd $m$ will appear around momentum $\pi$. (This is a standard side-effect of lattice regularization.)

\begin{figure}[htbc]
	\centering
	\includegraphics[width=\columnwidth]{{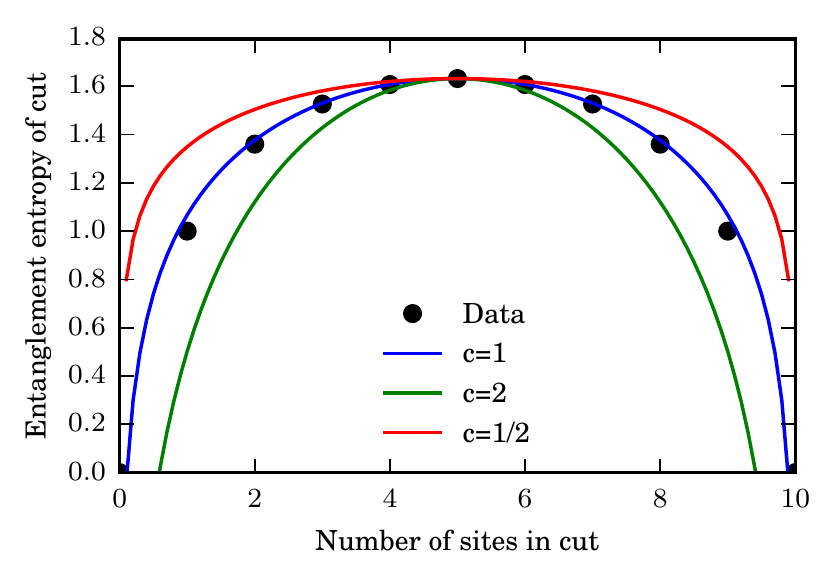}}
	\caption{Entanglement entropy within the entanglement ground state
of the soft-core boson state on $10$ sites. For comparison, the
Calabrese-Cardy formula~\cite{calabrese2004} $S(x) = c/3 \log \sin( \pi x/L) + const.$ is
shown with $c=\frac{1}{2}, 1,$ and $2$, with the $const.$ fixed by
matching the maximum of the entanglement entropy data. $c=1$ is a good
fit.}
	\label{fig:EdgeGS_EE}
\end{figure}

Given the PEPS representation, we can express the entanglement Hamiltonian $H_L$ for
the left semi-infinite cylinder, defined via $\rho_L = \exp(-H_L)$, as a Hamiltonian acting on the auxiliary
degrees of freedom of the PEPS crossing the cut, which we have denoted as $|\sigma_1,\ldots,\sigma_W)$.
We expect this Hamiltonian to encode the universal properties of the edge CFT, which should be invariant
under local gauge choices in the PEPS. Its ground state is (up to normalization) given as
\begin{equation}
|\Psi_0) = \sum_{\sigma_1,\ldots,\sigma_W} \langle \Psi^{(\sigma_1,\ldots,\sigma_W)}|\Psi_L^{(0)}\rangle |\sigma_1,\ldots,\sigma_W).
\end{equation}
In Fig.~\ref{fig:EdgeGS_EE}, we show the bipartite von Neumann entanglement entropy of this state for a cut
into $l$ and $W-l$ sites, which confirms the central charge $c=1$ of the edge CFT. A similar construction was
considered in \cite{Lou2011}.

\section{Symmetry protection}
\label{sec:symmetry}

\subsection{Overview}

While the gapless entanglement spectrum observed above is consistent with a symmetry-protected
topological phase, it does not by itself guarantee the presence of such a robust phase, and does not
allow us to infer which symmetries are protecting the topological properties of the phase.
A key observation that allows us to make progress on these crucial questions is that many points
in the entanglement spectrum are degenerate. In particular, we find that for cylinders of odd circumference,
the entire spectrum is doubly degenerate.
In this section, we will discuss how
the corresponding degenerate Schmidt states are related through the action of a symmetry of the HFBI wavefunction.
As discussed in Ref.~\onlinecite{pollmann2010} and reviewed in the Appendix~\ref{Appendix:MPS},
this symmetry action can be used to diagnose one-dimensional symmetry protected topological order,
for which the degeneracy throughout the entire entanglement spectrum is a robust feature.
We will demonstrate that the odd circumference cylinders, considered as quasi-one-dimensional states,
are indeed SPTs protected by a combination of lattice inversion and charge parity symmetries.

While the Schmidt eigenstates are uniquely defined for non-degenerate eigenvalues of the reduced
density matrix, they are not unique when the spectrum is degenerate and any choice of orthonormal
states in the degenerate subspace represents a valid choice of Schmidt states. Applying
a unitary transformation $V^{ji}$, which respects $\sum_i V^{ji} (V^{ki})^* = \delta_{jk}$, on the
left Schmidt states must be accompanied by an appropriate transformation $(V^{ji})^*$ applied to
the right Schmidt states.

In particular, this allows the action of an on-site symmetry (or more generally,
any symmetry which commutes separately with the reduced density matrices
for the left and right half) to mix Schmidt states corresponding to degenerate eigenvalues.
The action of such a symmetry operator $U_g$ takes the form
\beq
\label{eq:symschmidt}
\begin{split}
U_g \ket{\psi^{(i)}_{L}} &= \sum\limits_j \ket{\psi^{(j)}_{L}} V_g^{ji} \\
U_g \ket{\psi^{(i)}_{R}} &= \sum\limits_k \ket{\psi^{(k)}_{R}} \left(V_g^{ki} \right)^*,
\end{split}
\eeq
where the $V_g^{ji}$ are unitary matrices that only act on degenerate blocks of Schmidt states.
Crucially, Ref.~\onlinecite{pollmann2010} describes a numerical procedure to calculate $V_g$ for
an on-site symmetry $g$ within the MPS formalism, which we review in Appendix~\ref{Appendix:MPS}.

We can also analyze the effects of symmetries that preserve the entanglement cut but swap
the left and right halves of the cylinders. In general, we will consider
any symmetry $h$ that swaps the cylinder sides and squares to the identity, which we will call an
inverting symmetry. These satisfy a modification of~\eqnref{eq:symschmidt}:
\beq
\label{eq:isymschmidt}
\begin{split}
U_{h} \ket{\psi^{(i)}_{L}} &= \sum\limits_j \ket{\psi^{(j)}_{R}} V_{h}^{ji} \\
U_{h} \ket{\psi^{(i)}_{R}} &= \sum\limits_k \ket{\psi^{(k)}_{L}} \left( V_{h}^{ki} \right)^*.
\end{split}
\eeq
Note here that the left and right Schmidt states are exchanged in the transformation. We can
introduce a map $S$ that acts as
\beq \label{eq:S}
S \ket{\psi^{(i)}_{R}} = \ket{\psi^{(i)}_{L}}.
\eeq
Since a change in phase $\ket{\psi^{(i)}_{R}} \to e^{i \varphi} \ket{\psi^{(i)}_{R}}$ must be
accompanied by the complex conjugate $\ket{\psi^{(i)}_{L}} \to e^{-i \varphi} \ket{\psi^{(i)}_{L}}$
to preserve the Schmidt decomposition, $S$ is antiunitary.

Combining the above, we see that
\beq
\label{eq:isymschmidt2}
U_h S \ket{\psi^{(i)}_{R}} = \sum\limits_j \ket{\psi^{(j)}_{R}} V_{h}^{ji}
\eeq
defines the action of the operator $U_h S$ on the right Schmidt states (of course an equivalent
action can be defined on the left Schmidt states). Since $S$ is anti-unitary, the combined action
of $U_h S$ is also anti-unitary.
Together with the requirement that the symmetry squares to the identity, one finds that
(where $\mathbf{K}$ represents complex conjugation in the canonical basis)
\begin{equation}
\label{eq:antiunitarysym}
V_h V_h^* = (V_h \mathbf{K})^2 = e^{i \phi_h} I = \pm I,
\end{equation}
that is the inverting symmetry forms an anti-unitary projective representation of $\mathbb{Z}_2$.

As reviewed in Appendix~\ref{Appendix:MPS}, the collection of $V_g$ for on-site symmetries
sometimes fail to satisfy the group multiplication laws, i.e. one may find $V_{g_1 g_2} \neq
V_{g_1} V_{g_2}$.
Instead, they may form a projective representation, where group multiplication laws are obeyed up
to phases $\omega(g_1, g_2)$, i.e. $V_{g_1} V_{g_2} = \omega(g_1, g_2) V_{g_1 g_2}$.
Certain combinations of these phases, such as
\beq
e^{i\phi_{g_1, g_2}} \equiv \frac{\omega(g_1, g_2)}{\omega(g_2, g_1)}
\eeq
whenever $[g_1, g_2] = 0$, are \em symmetry protected topological invariants\em, which take
discrete values and hence cannot be changed continuously.
Thus, $\phi_{g_1, g_2} \neq 0$ signifies that the entanglement degeneracy cannot be removed without
breaking the symmetry or going through a phase transition.
Similarly, for the inverting (anti-unitary) symmetries $h$, the phase $\phi_h = \pi$ in
\eqnref{eq:antiunitarysym} signifies that
the degeneracy cannot be removed without breaking the symmetry~\cite{pollmann2010}.

\subsection{Symmetry protection of the HFBI}

The on-site symmetries of the featureless boson insulator considered here
are the \uone{} charge symmetry and the
anti-unitary time-reversal symmetry $\tau$, which acts by complex conjugation in the boson number basis.
Despite being at half-filling, the hard-core boson variant of the state does not have a particle-hole
symmetry. Exploring the edge action of these symmetries numerically, we find that they are all
represented linearly and thus do not protect the degeneracy of the entanglement spectrum
on cylinders of odd circumference.
In order to protect the degeneracy, we must therefore include lattice symmetries.

\begin{figure}
  \includegraphics[width=0.8\columnwidth]{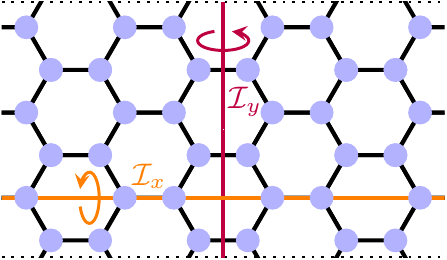}
  \caption{Lattice symmetries considered here:
  (i) $\I_x$ reflection about a line parallel to the long direction of the cylinder,
  (ii) $\I_y$ reflection about a line perpendicular to the long direction, corresponding to the entanglement
  cut shown in Fig.~\ref{fig:FBI_PEPS_2}. These are both chosen such that the reflection line
  crosses the hexagon center. Their product, $\I = \I_x \I_y$, thus represents (iii) the inversion about
  a hexagon center. \label{fig:symm} }
\end{figure}

By choosing a cylinder geometry, we explicitly break some of the lattice rotational
and reflection symmetries. The remaining symmetries are generated by
translations $T_x$ parallel and $T_y$ perpendicular to the cylinder axis
as well as reflections $\I_x$ about a line parallel and $\I_y$ about a line perpendicular to
the cylinder axis. We also consider lattice inversion $\I = \I_x \I_y$, equivalent to a $\pi$
rotation of the spatial plane about the center of a hexagonal plaquette. These symmetries are
illustrated in Fig.~\ref{fig:symm}.
We find that a number of symmetry-protected topological invariants that are defined through these symmetries
take non-trivial values in the HFBI, thus protecting the doubly degenerate entanglement spectrum
on odd circumference cylinders. The complete list of non-trivial invariants is summarized in
Table~\ref{table:sym}.

The crucial ingredient underlying these SPT invariants is a spatial symmetry $h$ that swaps the
two sides of the entanglement cut.
By a general symmetry analysis of \eqnref{eq:isymschmidt2}, $V_h$ must act as a
particle-hole symmetry on the edge, since the Schmidt pairing $S$ (\eqnref{eq:S}) pairs states with
opposite quantum numbers. In this case, the symmetry action $V_{\I_y}$ is precisely
that of a particle-hole transformation in the local PEPS basis.
Defining $\vket{\vec{\sigma}}=\vket{\sigma_1,\ldots,\sigma_W}$
and $\vket{1-\vec{\sigma}}=\vket{1-\sigma_1,\ldots,1-\sigma_W}$, we find that
\begin{equation}
V_{\I_y}\vket{\vec{\sigma}} = \vket{1-\vec{\sigma}},
\end{equation}
since a state where the $i^{th}$ hexagon contributes $\sigma_i$ bosons on the right is paired with
a state where the $i^{th}$ hexagon contributes $1-\sigma_i$ on the left. We can thus read off
that $V_{\I_y}$ acts like $\sigma_x$ in the space spanned by the states
$\lbrace \vket{\vec{\sigma}}, \vket{1-\vec{\sigma}} \rbrace$.

When $W$ is odd, these states have opposite charge parity. Specifically, if $\varPi = e^{i \pi \mathcal{Q}} \in U(1)$
is the charge parity symmetry, we have
\begin{align}
V_{\varPi} \vket{\vec{\sigma}} &= (-1)^{\sum \sigma_i} \vket{\vec{\sigma}} \nonumber\\
V_{\varPi} \vket{1-\vec{\sigma}} &= (-1)^{\sum (1-\sigma_i) } \vket{1-\vec{\sigma}} \nonumber\\
&= (-1)^W (-1)^{\sum \sigma_i} \vket{1-\vec{\sigma}}
\end{align}
Therefore, for $W$ odd, $V_{\varPi}$ acts like $\sigma_z$ in the space $\lbrace \vket{\vec{\sigma}}, \vket{1-\vec{\sigma}} \rbrace$.
It is thus reasonable to expect that the combination of these two symmetries acts as $V_{\varPi \I} = \sigma_x \sigma_z$,
which would obey the property that $V_{\varPi \I} V_{\varPi \I}^* = -I$ and thus form a topological invariant.

\begin{table}
\begin{tabular*}{\columnwidth}{@{\extracolsep{\stretch{1}}}*{5}{r}@{}}
\toprule
Group & Generators & Invariant & $i$ \\
\midrule
$\mathbb{Z}_2^P$ & $\{\varPi \I \}$
& $V_{\varPi \I} V_{\varPi \I}^* = -I$ &$-$  \\
$\mathbb{Z}_2^P$ & $\{\varPi \I_y \}$
&$V_{\varPi \I_y} V_{\varPi \I_y}^* = -I$ &$-$ \\ \hline
$\mathbb{Z}_2 \times \mathbb{Z}_2^{PT}$& $\{\varPi, \tau \I\}$
&$V_{\varPi} V_{\tau \I} V_{\varPi}^{-1} V_{\tau \I}^{-1} = -I$ &$+$ \\
$\mathbb{Z}_2 \times \mathbb{Z}_2^{PT}$& $\{\varPi, \tau \I_y\}$
&$V_{\varPi} V_{\tau \I_y} V_{\varPi}^{-1} V_{\tau \I_y}^{-1} = -I$ &$+$ \\
$\mathbb{Z}_2 \times \mathbb{Z}_2^{PT}$& $\{\varPi \I_x, \tau \I\}$
&$V_{\varPi \I_x} V_{\tau \I} V_{\varPi \I_x}^{-1} V_{\tau \I}^{-1} = -I$ &$+$ \\
$\mathbb{Z}_2 \times \mathbb{Z}_2^{PT}$& $\{\varPi \I_x, \tau \I_y\}$
&$V_{\varPi \I_x} V_{\tau \I_y} V_{\varPi \I_x}^{-1} V_{\tau \I_y}^{-1} = -I$ &$+$ \\
\bottomrule
\end{tabular*}
\caption{Summary of symmetry protecting invariants found for the HFBI state. The last column indicates
whether the symmetry acts unitarily ($i=+$) or antiunitarily ($i=-1$) on the edge.
The degenerate entanglement spectrum cannot be split unless all 6 of the  minimal protecting symmetry groups are broken.
\label{table:sym} }
\end{table}

The local PEPS basis is not unitarily equivalent to the canonical form basis, so we must check this numerically by performing an explicit calculation in the canonical form of an MPS representation of the state, as outlined in Appendix~\ref{Appendix:MPS}.
We thus confirm SPT invariants for symmetries that involve such a spatial symmetry $h$ and
an on-site symmetry.
There are several appropriate invariants, as listed in Table~\ref{table:sym}; the simplest is
\begin{equation}
V_{\varPi \I} V_{\varPi \I}^* = -I.
\end{equation}
From this we see that the charge, translation, and inversion symmetry can all
be broken without splitting the entanglement degeneracy, as long as the single combined symmetry
$\varPi \I $ is preserved. In Section~\ref{sec:perturbations}, we will discuss perturbations that preserve
this symmetry.

We note that there are also symmetries that act unitarily on the edge and yield SPT invariants;
however, these must form the group $\mathbb{Z}_2 \times \mathbb{Z}_2$ as $\mathbb{Z}_2$
does not have unitary projective representations. Examples for this are formed by involving
time-reversal symmetry; since $V_{\tau}$ and $V_{\I}$ both act antiunitarily, $V_{\tau \I}$ acts
unitarily on the edge. The $\mathbb{Z}_2 \times \mathbb{Z}_2$ group generated by
$\tau \I$ and $\varPi$ has a projective representation characterized by the
topological invariant
\beq
V_{\varPi} V_{\tau \I} V_{\varPi}^{-1} V_{\tau \I}^{-1}
 = - I.
\eeq
This symmetry protection gives a distinct class of perturbations that cannot
split the entanglement degeneracy.
The complete set of symmetry groups we find is summarized in Table~\ref{table:sym}.

We can form variants of the HFBI state, which are unitarily related to the original
state by an on-site unitary and thus share the same entanglement spectrum, where
the entanglement degeneracy can be protected by a lattice symmetry alone without involving
the on-site $\varPi$ symmetry.
Essentially, the protecting symmetries of the variant generated by a unitary $U$ is obtained by conjugating the generators of the protecting symmetries of the HFBI by $U$. These will be discused further in Appendix~\ref{Appendix:Variants}.

We also mention that the symmetry protected invariants produced above imply the existance of non-local correlations in the form of `membrane' order parameters that naturally generalize the string order parameters
from one-dimensional SPT phases \cite{pollmannturner2012}. For example, the sign of $V_{\varPi \I} V_{\varPi\I}^*$ can be detected by measuring the overlap of the state $\ket{\psi}$ and the same state with a partial application of the protecting symmetry, i.e.
\begin{equation}
\lim_{n\to\infty} \braopket{\psi}{(\varPi \I)_{1, 2n}}{ \psi} \propto (-1)^W,
\label{eq:membrane}
\end{equation}
where $(\varPi \I)_{1, 2n}$ is the restriction of $\varPi \I$ to $2 n$ cylinder slices.
We leave open the question of whether this `membrane' order parameter generalizes in any way to regions that do not wrap the cylinder.

\subsection{Tight-Binding Restriction}
Before concluding this section, we comment briefly on the role played by the restriction to a particular tight-binding model. 
Restricting to a particular tight-binding model is a stronger condition than merely specifying a space group symmetry. For instance, the triangular, honeycomb, and kagome lattices all share the same space group, but encode it using one, two and three orbitals per unit cell, respectively.   An example is graphene: the electronic Dirac cones in graphene are protected (for vanishingly small spin-orbit coupling) not solely by its space group symmetries, but rather by the tight-binding representation of those symmetries~\cite{RevModPhys.81.109}~\footnote{For additional discussion of this point, see~\onlinecite{parameswaran2013}}. A restriction to a tight-binding representation is often well motivated by experiments.
Choosing a tight-binding model amounts to restricting  to the class of models that can be represented using precisely the orbitals we began with; if we are given the freedom to {\it add} sites, it may be possible to exit a topological phase and enter a trivial one without closing the gap, while still preserving lattice symmetry. For the HFBI, this would be accomplished by adding sites at hexagon centers, and adiabatically deforming the relative weights afforded to bosons placed at the new sites and the original honeycomb lattice positions. The symmetry protection of the entanglement structure is thus fairly subtle. 

This subtlety is better understood for the simpler case of non-interacting fermions. The classification of free-fermion topological phases is known to be richer if one removes the freedom to add trivial bands~\cite{kitaev2009}. The continuity between two phases which arises upon the addition of such trivial bands is known as ``stable equivalence'' in accord with a basic notion in K-theory~\cite{kitaev2009}. 
We note that this tight-binding restriction also distinguishes the HFBI and related symmetry-protected short-range entangled states from the  ground states of ``filling-enforced'' topological band insulators introduced very recently \cite{Po2015}. 


\section{Quasi-local parent Hamiltonian and perturbations}
\label{sec:perturbations}

We now re-examine the question of whether the HFBI state is representative of a robust phase
of matter that is separated from conventional phases by phase transitions.
One way to demonstrate this -- beyond the topological invariants discussed above -- is to find a local Hamiltonian with a unique ground state that is the
HFBI wavefunction and study the ground state properties under perturbations to this Hamiltonian.
For many tensor network states -- those that satisfy an injectivity condition \cite{Perez-Garcia2008} --- a frustration-free, local parent Hamiltonian
with a unique ground state can be explicitly constructed. In our case, this injectivity condition can be shown to not hold on any simply connected
cluster of sites that we can numerically access, and this specific construction of a parent Hamiltonian is thus not possible.
Given the challenges of numerical simulations of two-dimensional systems, an exhaustive numerical search for such
a Hamiltonian seems unfeasible.

To avoid these problems, we will focus on a quasi-1D approach in this section. This is based on the observation that
while the PEPS is not injective on simply connected clusters, it does turn out to be injective on slices of an infinite
cylinder. This gives rise to a gapped Hamiltonian whose unique ground state is the HFBI. This `parent Hamiltonian' is local in the non-compact direction of the cylinder, but non-local around the cylinder and dependent on the circumference $W$.
We believe that nevertheless, the insights gained from these (partially non-local) Hamiltonians can serve as a starting point for
identifying the phase in more sophisticated numerical studies of fully two-dimensional boson systems.

Given the unperturbed Hamiltonian, we study the robustness of the entanglement spectrum to
perturbations. This depends on the class of perturbations allowed --
SPT phases are only distinct if perturbations that break the symmetry are forbidden, which is reflected in the
fact that the topological invariants that distinguish the phases are ill-defined in the absence of the symmetry.
According to the results discussed in Section~\ref{sec:symmetry}, it is not
necessary for the perturbations to preserve the entire symmetry group of the HFBI wavefunction
to preserve the entanglement in the state.
Instead, the entanglement is robust to any perturbation that does not break \em all \em of the six
protecting groups discussed in Table~\ref{table:sym}. This set of perturbations is much bigger than
the set of perturbations that preserve the entire symmetry group of the HFBI wavefunction. We
will confirm that the double degeneracy throughout the entire entanglement spectrum survives
these perturbations for odd-$W$ cylinders, while it splits for other perturbations that break all of
the protecting symmetries.


\subsection{Parent Hamiltonians for the $W=1$ cylinder and equivalence to the Haldane insulator}
The $W=1$ cylinder with hard-core bosons has a Hilbert space equivalent to a
two-leg spin-$\frac12$ ladder. The HFBI state in this case has a natural MPS representation
of bond dimension $d=2$, constructed by contracting the tensors around each cylinder slice.
A well-known property of MPS is the existence of a parent Hamiltonian -- a frustration-free
Hamiltonian with its unique ground state given by the MPS, first introduced by Ref.~\onlinecite{fannes1992}.
The parent Hamiltonian is constructed in this case as a translationally invariant sum of projectors,
where each term projects the Hilbert space of two consecutive rungs of the ladder to the $d^2=4$
dimensional subspace of states form spanned by the nonzero eigenvectors of the reduced density matrix
of those two rungs.
The result $H_0$ of this construction involves all possible terms that act on two rungs of the ladder and preserve charge
and reflection symmetry.

Using a local unitary transformation discussed in detail in Appendix~\ref{Appendix:aklttohfbi}, we
can transform the wavefunction of the HFBI on the $W=1$ cylinder to that of the `Haldane insulator'~\cite{berg2008,pollmann2010},
which is known to be the ground state of an extended Bose-Hubbard model on the two-leg ladder
in an appropriate parameter regime, and has also been shown to be a 1D SPT with a doubly
degenerate entanglement spectrum and a non-local string order parameter
protected by charge parity times inversion $\varPi \I$.

This extended Bose-Hubbard Hamiltonian that gives rise to the Haldane insulator includes only hopping and density-density
interactions; additionally, the range of these interactions extends only to neighboring rungs
of the ladder. It is thus clear that the additional interactions present in the parent Hamiltonian
for the HFBI can be tuned away without undergoing a phase transition.
The hard-core bosons should additionally be considered to have infinite on-site density
interactions, which can be tuned away from infinity to make a state with soft-core bosons.

\subsection{Perturbing the state on the $W=3$ cylinder}

Similar to the $W=1$ cylinder, we can obtain a parent Hamiltonian for the $W=3$ cylinder as a sum
of local projectors acting on adjacent slices of the cylinder. We then consider two different perturbations
to these quasi-local parent Hamiltonians.
For each perturbation, we use infinite time-evolving block decimation (iTEBD)~\cite{vidal2003,vidal2004,vidal2007classical,orus2007} to evolve
an initial wavefunction in imaginary time until it converges to the ground state of the perturbed Hamiltonian. The two perturbations considered are the superfluid pairing
\begin{equation}
H' = \Delta \sum\limits_{\ev{ij}} b_i b_j + h.c.,
\label{eq:pairing}
\end{equation}
which breaks the \uone{} charge symmetry down to the $\mathbb{Z}_2$ charge parity subgroup
and the uniform field
\begin{equation}
H' = h \sum\limits_{i} \left(b_i + b_i^{\dagger}\right),
\label{eq:field}
\end{equation}
which fully breaks \uone{} charge symmetry but preserves lattice symmetry.
The perturbation in \eqref{eq:pairing} 
does not break the protecting
symmetry $\varPi \I$, while the perturbation in \eqref{eq:field} breaks all of the protecting symmetries.

\begin{figure}[htbp]
	\centering
		\includegraphics[width=\columnwidth]{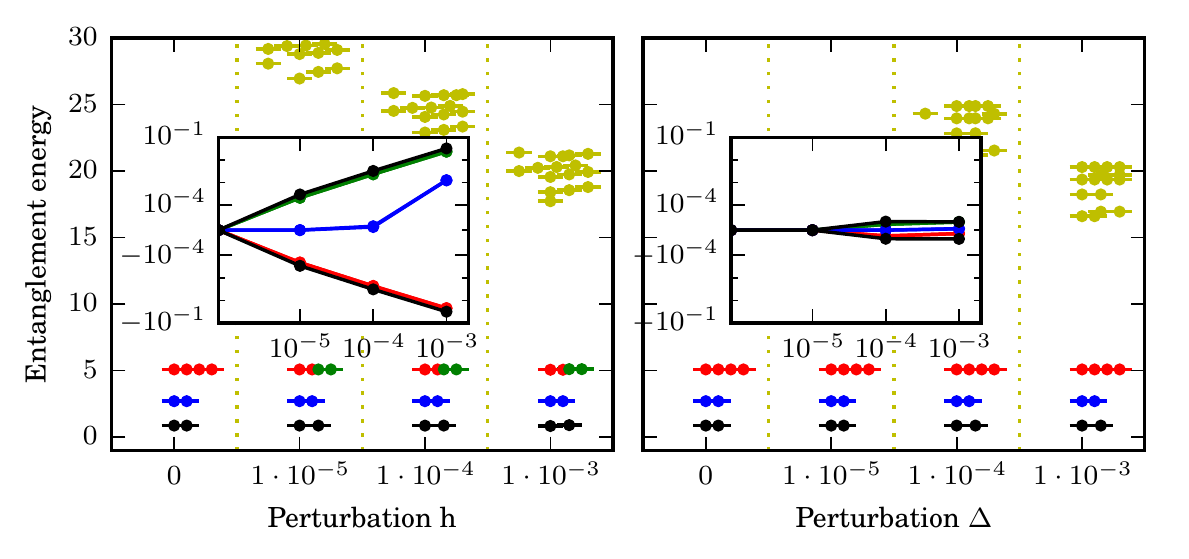}
	\caption{Entanglement spectrum for the ground state of the parent Hamiltonian on the $W=3$ cylinder under a symmetry-breaking perturbation~\eqref{eq:field} (left panel) and a symmetry-preserving perturbation~\eqref{eq:pairing} (right panel).
    These results obtained using an iTEBD simulation using a bond dimension of $M=24$.}
	\label{fig:perturbations}
\end{figure}

Figure~\ref{fig:perturbations} show the resulting
entanglement spectra from the ground states obtained with iTEBD. Indeed, the perturbation
in \eqref{eq:field} splits the degenerate entanglement spectrum, whereas the double-degeneracy
of the entire spectrum is preserved for those perturbations that do not break all of the
protecting symmetries. In the case of a symmetry-breaking perturbation, the splitting
is most easily observed for the higher levels, but -- as shown in the inset -- also
the lowest values of the entanglement spectrum are weakly split by an amount that
scales roughly linearly in the strength of the perturbation.

Unfortunately it is beyond the scope of this work to determine which perturbations leave the CFT structure of the entanglement spectra intact.
To assess this would require a Hamiltonian that is local in two dimensions (rather than the Hamiltonians used here which are only local
in the non-compact direction of the cylinder). Furthermore, it would require being able to perform accurate simulations for large
cylinders, which is prohibitive with the techniques used here.


\section{Boson-Vortex Duality}
\label{sec:duality}
An alternative approach for examining the phase structure of two dimensional bosonic systems is to use the boson-vortex duality, which rewrites the theory in terms of the superfluid vortex degrees of freedom defined on the dual lattice.
In this picture, the site filling of bosons on the original lattice is  mapped to an effective magnetic flux through dual lattice plaquettes that modifies the vortex hopping via the usual Aharonov-Bohm phases ~\cite{DasguptaHalperinDuality,FisherLeeDuality}. In the dual description the superfluid and Mott insulating phases of the bosons are respectively mapped into  the gapped and condensed phases of the vortices.  It is instructive to see how this approach fares on the honeycomb lattice at half-site-filling. The vortices move on the dual triangular lattice, and the original site filling of $1/2$ corresponds to a $\pi$-flux for vortices for every triangular lattice plaquette. Each unit cell on the triangular lattice contains a pair of triangles and hence $2\pi$ flux; as a consequence, the vortices transform normally (rather than projectively) under lattice symmetries. Naively, the $\pi$-flux has the effect of inverting the vortex band structure so that the vortex minima are shifted to the Brillouin zone corners $K$, $K'$ rather than the zone center  $\Gamma$. Condensing vortices at the zone corners would break lattice symmetries~\cite{Zaletel}. However, the fact that the vortices transform regularly under lattice symmetries --- in other words, that the flux pattern does not lead to an enlarged magnetic unit cell --- allows us to add additional hopping while preserving symmetries to returns the vortex minimum to $\Gamma$. Condensing vortices at $\Gamma$ then restores the $U(1)$ symmetry while preserving all lattice symmetries. As there are no other known symmetry-preserving insulating phases of bosons on the honeycomb lattice at this filling, we conjecture that this phase is adiabatically connected to the HFBI. 

It is perhaps worth noting that the vortex-condensation picture also  illustrates a fundamental distinction between half-filling on Bravais and symmorphic non-Bravais lattices. As an example, consider half-filling on the square lattice~\cite{PhysRevB.71.144508}; performing the duality transformation, we arrive at a theory of vortices moving on the dual square lattice with $\pi$ flux through each plaquette. Crucially, this flux assignment on the square lattice doubles the unit cell, and so vortices form a  projective representation of the space group (related to the magnetic translations familiar from studying particles in a magnetic field). This projective structure cannot be removed and so {\it guarantees} that a single non-degenerate minimum cannot be restored for any choice of vortex hopping parameters. Put differently, the vortices are (unlike in the honeycomb case) forced to carry non-trivial space group quantum numbers, and the condensation of single vortices necessarily breaks the symmetry~\cite{PhysRevB.71.144508}. Other approaches lead to more exotic alternatives, e.g. condensing vortices in pairs triggers fractionalization. This in accord with the expectation that a gapped featureless insulator is absent on the half-filled square lattice~\cite{Oshikawa:2000p1,PhysRevLett.90.236401,Hastings2004,Hastings:2005p1}. 


\section{Concluding Remarks and Discussion}

We have applied recently developed tensor network methods to study the edge properties
of a bosonic insulator that is featureless in the bulk. Our simulations are performed for an infinitely
long cylinder of finite circumference $W$. This allows us to numerically extract the exact entanglement
spectrum for up to $W = 10$. We find that the entanglement gap closes as $1/W$, and that furthermore
the low-lying spectrum coincides to high accuracy with the spectrum of a free boson conformal field theory.
This is further corroborated by observing a central charge of $c=1$ in the entropy of the lowest Schmidt state.

While these observations are consistent with and strongly suggestive of a symmetry-protected topological
phase, where such a gapless spectrum would naturally emerge at the edge, these calculations do not establish
a rigorous connection between the edge spectrum and symmetry-protection, i.e. they leave open the possibility
that the gapless entanglement spectrum is accidental.
To make progress on this question, we analyze in some detail the exact degeneracies in the entanglement
spectrum for cylinders of odd circumference $W$. Using recently developed tools based on matrix-product
states, we are able to establish a strong connection to the symmetries of the state by computing topological
invariants that detect the non-trivial action of certain symmetry operations. These symmetry operations, whose action is non-trivial, consist of particular combinations of lattice and spin symmetries on the edge.
This establishes in the affirmative that the quasi-one-dimensional systems obtained for odd cylinder widths $W$ represent
one-dimensional symmetry-protected topological phases.

We cannot establish with the same rigor that the symmetries that protect these one-dimensional topological
invariants also protect the gapless edge spectrum on the edge of the two-dimensional system. However, several
considerations are in favor of this.
Firstly, we observe that the symmetries that are shown to be relevant to the case of odd $W$ are not
inherently one-dimensional and could apply equally well to the full, two-dimensional system. The partial application of symmetry in the non-local order parameter in Eq.~\eqref{eq:membrane} could be applied to arbitrary inversion-symmetric regions in the plane, and not only to cylinder slices.
Additionally, we can construct an argument based on the picture of the edge physics provided by the tensor network representation.
As outlined in Section~\ref{sec:ES}, the edge of the tensor network representation with the cut chosen here
can be represented using the Hilbert space of a model of hard-core bosons hopping on a one-dimensional chain with one site
per plaquette, where the occupation of a site corresponds to whether the boson of that plaquette is found
on the left or right side of the cut. In this representation, the reflection symmetry about the cut takes
the special role of guaranteeing equal probability for the boson to be on the left or right of the cut, and thus fixing the model for the edge to half-filling. Thus, if the edge remains translationally symmetric, our model
for the edge has fractional charge per unit cell. If the entanglement Hamiltonian can be thought of as local, the Lieb-Schultz-Matthis theorem applies and guarantees that the entanglement edge is either gapless or spontaneously breaks a symmetry. This suggests that the phase is a two-dimensional symmetry-protected topological phase with a protecting group that includes translation and $\varPi \I$.

The calculations presented here provide a case study where tensor network representations lead to novel
insights into strongly correlated physics beyond what is accessible to more traditional methods, such as
the quantum-to-classical mappings pursued in Ref.~\onlinecite{kimchi2013} and reviewed in Section~\ref{sec:fbi}.
The tensor-network techniques used in the present approach allow us to strengthen the conclusions of
Ref.~\onlinecite{kimchi2013}, in particular on the absence of topological order, and reveal entanglement
properties that are entirely out of reach of quantum-to-classical mappings.
It is amusing to note that this development in theoretical methodology closely parallels the history of the prototypical SPT phase, the AKLT
phase of the spin-$1$ chain, where the existence of a quantum-to-classical mapping was known well before
the nontrivial entanglement structure was understood. As in that example, we expect that here
as well, the quantum-to-classical mappings are restricted to rather special points within a broader SPT phase,
whereas the tensor-network description and its corresponding entanglement structure are expected to be
valid more generally throughout the phase.

The question of a parent Hamiltonian, i.e. whether the HFBI can be established as the unique ground state of a gapped local Hamiltonian, remains open. As reviewed
briefly in Sec.~\ref{sec:perturbations}, the structure of the PEPS does not allow us to straightforwardly
extract a local parent Hamiltonian in two dimensions. However, this by no means implies that such a
parent Hamiltonian does not exist, and future work will explore different numerical approaches to find such a
Hamiltonian.

\noindent{\bf Note:} While completing this work, we became aware of related PEPS constructions of featureless paramagnetic wavefunctions on the square lattice with spin $1$ per site, and on the honeycomb lattice with spin $1$ or $\frac{1}{2}$ per site~\cite{jian2015}. The spin-$\frac{1}{2}$ honeycomb lattice example corresponds to the same filling as the featureless insulating phase considered here, but has higher symmetry ($SO(3)$) compared to the \uone{} symmetry in the present paper. We note that even in the case where we consider spinful fermions (see Appendix \ref{appendix:fermionic}) bound into Cooper pairs, the wavefunction we construct here is {\it not} a valid wavefunction for a spin-only model: projecting it to the case of single-fermion occupancy per site (as appropriate to a spin model) annihilates the wavefunction. It will be interesting to study if the spin-only wavefunctions constructed in Ref.~\onlinecite{jian2015} possess similarly rich entanglement structure as the HFBI.
\acknowledgments
We thank Meng Cheng, Dominic Else, Eugeniu Plamadeala and Ashvin Vishwanath for useful discussions.
I.K. and S.A.P. thank Ari Turner, Fa Wang, and Ashvin Vishwanath for collaboration on related work~\cite{kimchi2013}. We would also like to thank Michael Zaletel and Chao-Ming Jian for sharing the results of Ref.~\onlinecite{jian2015} prior to publication.
S.A.P. acknowledges support from the National Science Foundation under Grant No. DMR-1455366.

\appendix

\section{Determining the edge action of the symmetry using MPS}
\label{Appendix:MPS}

We can use the formalism of matrix-product states to determine the action
of physical symmetries on the Schmidt states. First, this will lead to the assignment
of charge and translation (which both act on-site in the MPS representation) quantum numbers
to the Schmidt states and corresponding entanglement spectrum as labeled in, e.g., Fig.~\ref{fig:ESL910}.
Secondly, this will be used to numerically extract the topological invariants discussed in
Section~\ref{sec:symmetry}.
We now review this formalism briefly, including a discussion of the method that allows us to numerically
determine the symmetry action of inversion symmetry on the Schmidt states.
Both of these discussions follow Ref.~\onlinecite{pollmann2010}.

We start by finding tensors $\Gamma$, $\Lambda$ representing the
so-called canonical form of the MPS, as detailed in Refs.~\onlinecite{vidal2003,vidal2007classical}:
\beq
\ket{\psi} = \sum\limits_{\{p_i\}} \ldots \Lambda \Gamma_{p_0} \Lambda \Gamma_{p_1} \Lambda \Gamma_{p_2} \Lambda \ldots \ket{... p_0 p_1 p_2 ...}.
\eeq
This canonical form provides the Schmidt decomposition at each site in the lattice.
Here, each physical leg of the MPS represents all $2W$ physical sites on a cylinder slice,
and each virtual leg represents all virtual indices that connect cylinder slices; the bond dimension
of the MPS is thus $2^W$.
The change of basis to canonical form generally mixes the Hilbert spaces from these
virtual legs, so the resulting basis won't be local around the circumference
of the cylinder.

For each on-site symmetry of the wavefunction $U_g = \otimes_i u^i_g$, with $U_g
\ket{\psi} = e^{i \Theta_g} \ket{\psi}$, there is an operator $V_g$ that
acts on the virtual leg of the MPS and satisfies the equation
\begin{center}
\beq
\label{eq:onsitesym}
\eeq
\vskip-5em
\includegraphics[width=0.6\columnwidth]{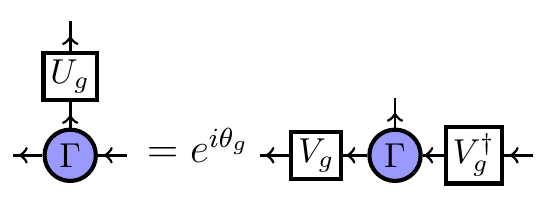}.
\end{center}
This equation can be rewritten and solved as an eigenvector problem;
for an MPS with a nondegenerate largest transfer matrix eigenvalue,
this equation is guaranteed have a unique solution where the eigenvalue $e^{i \theta_g}$
is the largest eigenvalue of the eigenvector problem.
These solutions $V_g$ have two important properties: they are only defined up to a phase,
and they are guaranteed to commute with the diagonal matrix $\Lambda$ of Schmidt weights.

Due to the first property, these operators are not guaranteed to obey the group multiplication
laws, i.e. one could find situations where
\begin{equation}
V_g V_h = \omega(g,h) V_{gh}.
\end{equation}
It is not always possible to absorb these phases into the definitions of the $V_g$; in those
cases, the $V_g$ do not form a linear representation of the group but rather a projective representation.
The set of equivalent classes of phases $\omega(g, h)$
under redefinitions $V_g \to \alpha(g)V_g$ is called $H^2(G, U(1))$, the second group
cohomology with \uone{} coefficients.

For all the groups discussed in this paper, the group cohomology classes are labeled by
elements of a discrete abelian group. These discrete classes cannot be connected to each other
continuously without undergoing a bulk phase transition or breaking the symmetry.
Additionally, the classification of projective representations for the on-site symmetry group
$U(1) \times \mathbb{Z}_W$ representing charge and translation around the cylinder is trivial.
Thus, these edge symmetries can be taken to act linearly,
and all Schmidt states can always be simultaneously assigned charge and momentum eigenvalues,
as in Figure~\ref{fig:ESL910}.

The second property guarantees that the $V_g$ only mixes exactly degenerate Schmidt states.
The action of $V_g$ must have the same phases $\omega(g, h)$ on each degenerate block of Schmidt
states, so the projective representation can be nontrivial on any block only if every Schmidt
state throughout the entire spectrum is degenerate. The degeneracy will be protected by the
symmetry if and only if the $V_g$ form a nontrivial projective representation. Therefore
this 1D SPT analysis can only potentially give a nontrival answer for the odd $W$ states
of the HFBI, where this exact degeneracy is seen throughout the spectrum.

The MPS analysis of inversion symmetry proceeds similarly. We will consider in general any symmetry
$h$ of the wavefunction that squares to the identity and that can be written in the MPS as the
product of an on-site symmetry action $U_h$ and a transpose of the site tensor.
This will include an inversion of the honeycomb lattice - equivalent to a 180 degree rotation
about the center of any plaquette, which we label $\I = \I_y \I_x$, and the combination of
inversion with on-site symmetries. In addition, by blocking two site-tensors together, we
can write the reflection symmetry $\I_y$ in this form as well. In this scenario, the
edge symmetry action satisfies
\begin{center}
\beq
\label{eq:invsym}
\eeq
\vskip-5em
\includegraphics[width=0.6\columnwidth]{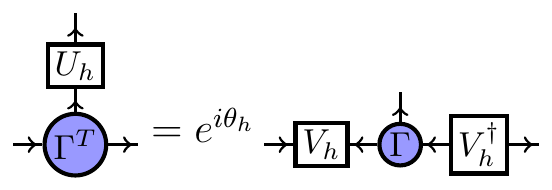}.
\end{center}
The map $V_{h}$ is also computed from an eigenvector problem.

For the HFBI, the symmetry group respected by the cylinder geometry is
$U(1) \times (\mathbb{Z}_W \rtimes \mathbb{Z}_2)
\times \mathbb{Z}_2^P \times \mathbb{Z}_2^T$, where the factors refer to charge
symmetry, translation around the cylinder, $\I_x$, $\I_y$, and $\tau$ respectively.
The $P$ and $T$ denote space-reversing and time-reversing symmetries, and signify the
antiunitary action on the Schmidt states.
Many of the non-trivial projective
representations of such a complicated group will remain projective when the symmetry is
restricted to a subgroup - in this case, the full symmetry is not needed to protect the
entanglement degeneracy. As shown in Table~\ref{table:sym}, the projective representation
corresponding to the HFBI state can indeed be protected by any one of a number of subgroups
of the full symmetry group, all involving inversions and charge parity.

The symmetry actions -- both on-site and inversion symmetries --
are computed in the Schmidt basis, but can be transformed
into the basis $\vket{\{\sigma_i\}}$ determined by the virtual legs of the PEPS in
Figure~\ref{fig:FBI_PEPS_2}.
In this case, the symmetry action $V_{\I_y}$ is precisely
a particle-hole symmetry in the local PEPS basis, with coefficients
$$
V_{\I_y}\vket{\sigma_1, \ldots, \sigma_{W}} = \vket{1-\sigma_1, \ldots, 1-\sigma_{W}},
$$
since a state where the $i^{th}$ hexagon contributes $\sigma_i$ bosons on the right is paired with
a state where the $i^{th}$ hexagon contributes $1-\sigma_i$ on the left.
Thus
$$
V_{\I_y} = \prod\limits_i \sigma_i^x K,
$$
where K is complex conjugation in the local PEPS basis, and $\sigma_i^x$ is the Pauli
operator acting on the $i^{th}$ site of the local PEPS basis.

Charge symmetry acts locally as well:
$$
e^{i \theta \mathcal{Q}}\vket{\sigma_1, \ldots, \sigma_{W}} = e^{i \theta \sum(\sigma_i-1/2)}\vket{\sigma_1, \ldots, \sigma_{W}}.
$$
In particular, charge parity $V_{\varPi} = e^{i \theta \mathcal{Q}}$ can be written as
$$V_{\varPi} = e^{i \pi \sum(\sigma_i - 1/2)} = \prod\limits_i \sigma_i^z.$$

The combined action of charge parity and reflection across the cut takes the form
$$
V_{\varPi \I_y} = \prod\limits_i \left(i \sigma_i^y \right) K,
$$
which is precisely the form that time-reversal acting on an ordinary spin-$\frac12$ chain takes.
When the circumference of the cylinder $W$ is odd, we see that
$$V_{\varPi \I_y}V_{\varPi \I_y}^{*} = -I.$$
The degeneracy of the entanglement spectrum can be seen as an application of Kramer's theorem.
Formally, this property is said to characterize the nontrivial projective representation
$$
H^2(\mathbb{Z}_2^P; U(1)) = \mathbb{Z}_2,
$$
and remains true while $\varPi \I_y$ is a symmetry and no phase transitions have occurred.

Time reversal symmetry acts as complex conjugation in the local PEPS basis $V_{\tau}=K$.
Translation and $\I_x$ act as permutations of the local PEPS basis:
\begin{equation*}
\begin{split}
V_{T}\vket{\sigma_1, \ldots, \sigma_{W}} &= \vket{\sigma_2, \ldots, \sigma_{W}, \sigma_{1}} \\
V_{\I_x}\vket{\sigma_1, \ldots, \sigma_{W}} &= \vket{\sigma_W, \ldots, \sigma_{1}}.
\end{split}
\end{equation*}
These symmetries can be combined with $V_{\varPi \I_y}$ to create the additional topological
invariants shown in Table~\ref{table:sym}. A non-trivial projective
representation in $$H^2(\mathbb{Z}_2 \times \mathbb{Z}_2; U(1)) = \mathbb{Z}_2$$
is created whenever two unitary symmetries that commute in the bulk satisfy
$$V_{g_1} V_{g_2} V_{g_1}^{-1} V_{g_2}^{-1} = -I.$$
Each new invariant is related to a new set of pertubations that can't break the entanglement
degeneracy.

\section{From the AKLT to the $W=1$ HFBI}
\label{Appendix:aklttohfbi}
The AKLT state $\ket{\psi_\mathrm{AKLT}}$ is a state of a spin-1 chain that has an exact representation as an MPS of bond dimension 2 using site tensors $A^p_{ij}$ related to the Pauli matrices~\cite{schollwock2011}. It is the exact ground state of the AKLT Hamiltonian
\begin{equation}
H_\mathrm{AKLT} = \sum\limits_j \vec{S}_j \cdot \vec{S}_{j+1} + \frac13 (\vec{S}_j \cdot \vec{S}_{j+1})^2,
\end{equation}
but it is known that the simpler Hamiltonian
\begin{equation}
H_\mathrm{AF} = \sum\limits_j \vec{S}_j \cdot \vec{S}_{j+1}
\end{equation}
is in the same phase, i.e. the AKLT state lies in the Haldane phase of the spin-1 Heisenberg chain.
By a series of transformations, we can find a representative MPS wavefunction $\ket{\psi_\mathrm{HI}}$
and a simple representative Hamiltonian that can be adiabatically connected to the
$W=1$ HFBI and its corresponding parent Hamiltonian.

By using the unitary operator
\begin{equation}
U(\pi) = \prod\limits_{j \text{ even}} e^{i \pi S^z_j}
\end{equation}
which flips the $x, y$ components of the spins on every other site, we create a wavefunction representative
of the Haldane insulator (HI)~\cite{berg2008} phase, which is protected by $U \I U^{\dagger} = \varPi \I$~\cite{pollmann2010}.
This phase is obtained as the ground state of the Hamiltonian
\begin{align}
H' &= U H_{AF} U^{\dagger} \\ &= \sum\limits_j\left( -\frac12(S^+_j S^-_{j+1} + h.c.) + S^z_j S^z_{j+1} \right).
\end{align}

Each spin-1 degree of freedom can be split into a pair of $S=1/2$ spins to make a state on
a spin-$\frac12$  ladder. An appropriate Hamiltonian can be found in terms of the new
spin variables $\vec{S}_{j, A/B}$ by adding a term to project out the spin-singlet component
of $\vec{S}_{j,A}+\vec{S}_{j,B}$. The spin-$\frac12$'s can then be treated as hard-core bosons. The Hamiltonian becomes
\begin{multline}
H_\mathrm{HI} = \sum\limits_j -\frac{t}{2}((b^{\dagger}_{jA} + b^{\dagger}_{jB}) (b_{j+1A} + b_{j+1B}) + h.c.)\\
 + V(n_{jA} + n_{jB} - 1) (n_{j+1A} + n_{j+1B} - 1) \\
 - \frac{J}{2}(b^{\dagger}_{jA} b_{jB} + h.c.) - J(n_{jA}-\frac12)(n_{jB}-\frac12),
 \label{eq:w1ham}
\end{multline}
where $t=1$, $V = 1$, and $J \rightarrow \infty$. The $J$ term projects the spin-singlet out of each rung,
and in practice only needs to be larger than all other relevant scales to drive the system into the
appropriate phase.

We can do the same transformations on the MPS $\ket{\psi_\mathrm{AKLT}}$ to obtain a new
MPS $\ket{\psi_\mathrm{HI}}$ with bond dimension 2 and site tensor $A'^p_{ij}$ that represents a
state in the phase of $H_\mathrm{HI}$ on the two-leg ladder. The site tensor $S^p_{ij}$ of the $W=1$
HFBI also has bond dimension 2 and represents a state
of hard-core bosons on the two-leg ladder. Numerically, these are represented by the (unnormalized)
site tensors
$$
A'^p =
\begin{cases}
     \left(\begin{array}{cc}
      0 & 0 \\
      2 & 0
      \end{array} \right) & p=(00) \\
      \left(\begin{array}{cc}
      1 & 0 \\
      0 & 1
      \end{array} \right) & p=(01) \\
      \left(\begin{array}{cc}
      1 & 0 \\
      0 & 1
      \end{array} \right) & p=(10) \\
      \left(\begin{array}{cc}
      0 & 2 \\
      0 & 0
      \end{array} \right) & p=(11)
\end{cases}
$$
and
$$
S^p =
\begin{cases}
     \left(\begin{array}{cc}
      0 & 0 \\
      1 & 0
      \end{array} \right) & p=(00) \\
      \left(\begin{array}{cc}
      2 & 0 \\
      0 & 1
      \end{array} \right) & p=(01) \\
      \left(\begin{array}{cc}
      1 & 0 \\
      0 & 2
      \end{array} \right) & p=(10) \\
      \left(\begin{array}{cc}
      0 & 5 \\
      0 & 0
      \end{array} \right) & p=(11),
\end{cases}
$$
where $p = (p_1 p_2)$ represents the occupation numbers of the hard-core bosons on the two sites on each leg of the ladder.

By linearly tuning the site tensors using
\begin{equation}
S^p_{ij}(t) = t A'^p_{ij} + (1-t) S^p_{ij},
\end{equation}
and checking that the transfer matrix of the resulting state is non-degenerate for all $t \in [0, 1]$,
we confirmed that the $W=1$ HFBI can be tuned in the space of bond dimension $d=2$ MPS to $\ket{\psi_\mathrm{HI}}$
without passing through a phase transition,
and the representative Hamiltonian in \eqref{eq:w1ham} describes a state in the same phase.

By calculating the canonical form of the $S^p_{ij}$ site tensor, one can check that the $W=1$ HFBI wavefunction has particle-hole symmetry, while the $W>1$ states do not. This particle-hole symmetry $C$ can also be used as a symmetry protection via the $\mathbb{Z}_2 \times \mathbb{Z}_2$ group generated by $\{\varPi, C\}$ or by the time-reversing symmetry $\varPi C \tau$. This fact is well known in the context of the AKLT state, where $\varPi$ and $C$ are represented in the spin-language as $\pi$ rotations about the $z$ and $x$ axes, and $\varPi C \tau$ is the time-reversing symmetry $i S_y K$ that flips all components of the spins.

In the context of the argument laid out in the conclusion, it seems that particle-hole symmetry can play the same role as inversion symmetry in ensuring the edge remains at half-filling.

\section{Variants on the HFBI wavefunction}
\label{Appendix:Variants}
\subsection{Tuning soft-core bosons to hard-core}
In Equations~\eqref{eqn:Dsc} and \eqref{eqn:Dhc}, the tensor $D$ can be replaced by a more general form
\begin{equation} \label{eqn:Dgen}
D_{p, i_0 i_1 i_2}  = \left\{ \begin{array}{ll}
													d_p  &: p =i_0+i_1+i_2  \\
													0  &:  \text{else}
													\end{array}
											\right. ,
\end{equation}
which the coefficients $d_p = 1,\, 1,\, \sqrt{2},\,\sqrt{6}$ for $p = 0,\, 1,\, 2,\, 3$ in the soft-core state and $d_p = 1,\, 1,\, 0,\, 0$ for $p = 0,\, 1,\, 2,\, 3$ in the hard-core state. We can continously tune the coefficients $d_2$ and $d_3$ from the soft-core to the hard-core values.
Upon doing so, we find that the transfer matrix spectrum remains gapped, with the correlation length monotonically increasing from the soft-core state to the hard-core state.
Furthermore, the low energy parts of the entanglement spectrum do not change significantly through this tuning.
Therefore we expect that the hard-core and soft-core phases can be adiabatically connected with a
path of local Hamiltonians, and all SPT results that apply to one state apply to the other. By choosing appropriate values of $d_2$ and $d_3$,
we can also make a state that is equivalent to replacing the vacuum $\ket{0}$ in
Equation~\eqref{eq:def} with a constant background of $N$ bosons on each site, $N \rightarrow \infty$,
and applying boson annihilation instead of creation operators. We can also make a state of spin-$S$ spins,
which is however not SU(2)-invariant, where Equation~\eqref{eq:def} becomes
\begin{equation} \label{eq:spindef}
\ket{\psi} = \prod\limits_{\hexagon} \left( \sum\limits_{i \in \hexagon} S^{+}_{i} \right) \prod\limits_i \ket{S^z_i = -S}.
\end{equation}
Here, the hard-core state would most closely correspond to a state of $S=1/2$ spins, while the soft-core state corresponds
to a state of $S=3/2$ spins. All of these states have the same symmetry protection properties.

\subsection{Interpretation as  a Fermionic Wavefunction}
\label{appendix:fermionic}
We can also interpret the hard-core variant of the HFBI as a wavefunction for spinful fermions on the honeycomb lattice at half filling. Note that including the spin, `full filling' of a site corresponds to a pair of fermions on each site, so half filling occurs with exactly one fermion per site, corresponding to two fermions per unit cell. Assuming no spin polarization, there must be an equal number of `up' and `down' spins. We can bind pairs of opposite-spin fermions into a Cooper pair, which yields one Cooper pair per unit cell. As a Cooper pair is equivalent to a hard-core boson, we may place the Cooper pairs into the hard-core variant of the HFBI. This is equivalent to the wavefunction
\begin{equation}
\ket{\Psi_{e}} = \prod_{\hexagon} \left(\sum_{i\in\hexagon} c^\dagger_{i\uparrow}c^\dagger_{i\downarrow}\right)\ket{0}.
\end{equation}
As the Cooper pair is in a spin singlet state, this wavefunction preserves $SU(2)$ spin symmetry, in addition to the lattice and \uone{} charge conservation symmetries. It is therefore a symmetry-preserving wavefunction of spinful fermions (i.e., electrons) on the honeycomb lattice at half filling. However, it is {\it not} a valid wavefunction for a pure $SU(2)$ symmetric spin model, as it has a vanishing projection onto the subspace where each site has exactly unit occupancy. Note that the necessity to have `preformed pairs' that can then be put into a hard-core boson state vividly illustrates the fundamentally interacting nature of this fermionic state.

\subsection{Inversion Protected Phase}
Additionally, the tensor $W$ in Equation~\eqref{eq:W} can be replaced by the more general form
\begin{equation} \label{eq:Wgen}
W^{n_1 \ldots n_6}  = \left\{ \begin{array}{lr}
													p_x  : & n_x=1,\, n_y = 0
													\; \forall \; y \neq x \\
													0  : & \text{else}
													\end{array} \right.,
\end{equation}
which corresponds to modifying Equation~\eqref{eq:def} to
\begin{equation} \label{eq:pdef}
\ket{\psi_{\ell}} = \prod\limits_{\hexagon} \left( \sum\limits_{i \in \hexagon} p_i b^{\dagger}_{i} \right) \ket{0}.
\end{equation}

This does not in general preserve the rotational symmetry of the state, but it does if the
coefficients $p_0, \ldots p_5$ are in an angular momentum mode $$p_x = e^{i x \ell}$$ where
$\ell \in \{0, 2\pi/6, \ldots,  5 \cdot 2\pi/6 \}$. These 6 discrete solutions can't be continously tuned to one another while preserving all the lattice symmetries.

The state $\ket{\psi_{\ell=\pi}}$ can be shown to be related to state $\ket{\psi_{\ell=0}}$ discussed in the main text by a on-site unitary operator $U(\pi)$,
where
\begin{equation} \label{eq:Uphi}
U(\varphi) = \prod\limits_{j \in B} e^{i \varphi \hat{Q}_j}.
\end{equation}
Due to this relation, $\ket{\psi_{\ell=\pi}}$ and $\ket{\psi_{\ell=0}}$ have identical correlation lengths and entanglement spectra.
However, the protecting symmetries from Table~\ref{table:sym} are mapped using conjugation
by $U(\pi)$ into a new set of protecting symmetries, shown in Table~\ref{table:pisym}.
Notably, since
\begin{equation}
U(\pi) \varPi \I U(\pi)^{\dagger} = \I,
\end{equation}
this state has doubly degenerate entanglement spectra on odd cylinder sizes protected by lattice inversion symmetry alone.
\begin{table}[t]
\vskip 1em
\begin{tabular*}{\columnwidth}{@{\extracolsep{\stretch{1}}}*{4}{r}@{}}
\toprule
Group & Generators & Invariant & $i$  \\
\midrule
$\mathbb{Z}_2^P$ & $\{\I \}$
& $V_{\I} V_{\I}^* = -I$ &$-$ \\
$\mathbb{Z}_2^P$ & $\{\I_y \}$
&$V_{\I_y} V_{\I_y}^* = -I$ &$-$ \\ \hline
$\mathbb{Z}_2 \times \mathbb{Z}_2^{PT}$& $\{\varPi, \tau \varPi \I\}$
&$V_{\varPi} V_{\tau \varPi \I} V_{\varPi}^{-1} V_{\tau \varPi \I}^{-1} = -I$ &$+$ \\
$\mathbb{Z}_2 \times \mathbb{Z}_2^{PT}$& $\{\varPi, \tau \varPi \I_y\}$
&$V_{\varPi} V_{\tau \varPi \I_y} V_{\varPi}^{-1} V_{\tau \varPi \I_y}^{-1} = -I$ &$+$ \\
$\mathbb{Z}_2 \times \mathbb{Z}_2^{PT}$& $\{\varPi \I_x, \tau \varPi \I\}$
&$V_{\varPi \I_x} V_{\tau \varPi \I} V_{\varPi \I_x}^{-1} V_{\tau \varPi \I}^{-1} = -I$&$+$ \\
$\mathbb{Z}_2 \times \mathbb{Z}_2^{PT}$& $\{\varPi \I_x, \tau \varPi \I_y\}$
&$V_{\varPi \I_x} V_{\tau \varPi \I_y} V_{\varPi \I_x}^{-1} V_{\tau \varPi \I_y}^{-1} = -I$ &$+$\\
\bottomrule
\end{tabular*}
\caption{Summary of symmetry protecting invariants found for the $\ket{\psi_{\ell=\pi}}$ state.
The degenerate entanglement spectrum cannot be split unless all 6 of the  minimal protecting symmetry groups are broken. }
\label{table:pisym}
\end{table}
Thus while the entanglement degeneracy in the HFBI state $\ket{\psi_{\ell=0}}$ is not split under
a staggered field
\begin{equation}
\begin{split}
H' = h^s \sum\limits_{i} (-1)^i \left(b_i + b_i^{\dagger}\right) \\ \text{ with } (-1)^i =
   \begin{cases}
      1 & i \in A \\
     -1 & i \in B
     \end{cases}
\end{split}
\label{eq:staggered}
\end{equation}
(which fully breaks \uone{} charge symmetry and inversion but not the combined symmetry $\varPi \I$),
the entanglement degeneracy in the state $\ket{\psi_{\ell=\pi}}$ would be unsplit by a uniform
field, which may be physically more interesting.

A similar mapping for 1-D inversion protected states is discussed in Appendix A of Ref.~\onlinecite{pollmann2010}. As discussed in Appendix~\ref{Appendix:aklttohfbi},
the state $\ket{\psi_{\ell=0}}$
on the $W=1$ cylinder is adiabatically connected to the 1-D Haldane insulator state~\cite{berg2008,pollmann2010}. Correspondingly,
the state $\ket{\psi_{\ell=\pi}}$ on the $W=1$ cylinder is adiabatically connected to the 1-D AKLT state.

\bibliography{fbi}

\end{document}